\pdfoutput=1
\documentclass[prd, nofootinbib, floatfix, notitlepage,twocolumn]{revtex4-1}

\usepackage{amsmath,amsfonts,amssymb}
\usepackage{mathrsfs}
\usepackage{graphicx}
\usepackage[english]{babel} 
\usepackage{hyperref} 
\usepackage{verbatim}
\usepackage[usenames, dvipsnames]{color}
\usepackage{bbold}
\usepackage{overpic}

\newcommand{\be}{\begin{equation}}
\newcommand{\ee}{\end{equation}}
\newcommand{\bes}{\begin{equation*}}
\newcommand{\ees}{\end{equation*}}

  \usepackage[utf8]{inputenc}

\begin{document}

\title{Spin Wave Analysis of Low-Temperature Thermal Hall Effect in the Candidate Kitaev Spin Liquid $\alpha$-RuCl${}_3$}

\author{Jonathan Cookmeyer$^{1}$}
\author{Joel E. Moore$^{1,2}$}
\affiliation{$^1$Department of Physics, University of California, Berkeley, CA, 94720, USA}
\affiliation{$^2$Materials Sciences Division, Lawrence Berkeley National Laboratory, Berkeley, California, 94720, USA}

\begin{abstract}
Proposed effective Hamiltonians from the literature for the material $\alpha$-RuCl${}_3$ are used to compute the magnon thermal Hall conductivity, $\kappa_{xy}$, using linear spin wave theory for the magnetically ordered state. No model previously proposed that was tested explains published experimental data. Models with Kitaev interaction $K>0$ are seen to predict $\kappa_{xy}\gtrsim 0$, which is inconsistent with the data. Fluctuations toward a Kitaev-type spin liquid would have the wrong sign to explain the data.  However, a slight variant of a previously proposed model predicts a large $\kappa_{xy}$, demonstrating that the low-temperature thermal Hall effect could be generated exclusively by the Berry curvature of the magnon bands. The experimental data of $\kappa_{xy}$ can therefore serve as another method to constrain a proposed effective Hamiltonian.
\end{abstract}

\date{\today}

\maketitle

\section{Introduction}
The Kitaev spin-spin interaction on a honeycomb lattice provides a solvable model with abelian and non-abelian anyonic excitations, as well as Majorana edge modes \cite{kitaev:2006anyons}.  Renewed interest in finding a material that realizes the Kitaev interaction was, in part, sparked by the observation that it can arise as an effective Hamiltonian for some transition metals with strong spin-orbit coupling \cite{jackeli:2009mott}.

The material $\alpha$-RuCl${}_3$ has generated much excitement due to it presenting some promise as such a Kitaev-magnet (see Ref.~\cite{trebst:2017kitaev} and references therein). The Ru atoms compose an ideal honeycomb lattice, and the material itself is a Mott insulator with the requisite spin-orbit coupling \cite{Koitzsch:2016jeff}. At low temperatures, the spins enter into a zigzag ordering, but a large in-honeycomb-plane field may induce a spin liquid state \cite{baek:2017evidence,banerjee:2017neutron,ran:2017spin,leahy:2017anomalous,kasahara:2018majorana}.

Raman scattering results on $\alpha$-RuCl${}_3$ reveal a continuum of excitations that exists above and below the magnetic transition temperature~\cite{Sandilands:2015Scattering}; after subtracting a bosonic background, the excitations appear to be fermionic, which suggests proximity to a spin liquid state~\cite{nasu:2016fermionic}. The continuum of excitations is also seen in inelastic neutron scattering data and seem to be qualitatively similar to that expected from the pure Kitaev model~\cite{banerjee:2016proximate,banerjee:2017neutron}, and linear spin wave theory fits to inelastic neutron scattering data suggest a significant Kitaev interaction~\cite{banerjee:2016proximate,ran:2017spin}. Experiments with terahertz spectroscopy \cite{Little:2017Antiferromagnetic,Wang:2017Magnetic,wu:2018field} and electron spin resonance \cite{Ponomaryov:2017Unconvetional} above and below the field induced transition further demonstrate the existence of interesting features in the excitation spectrum that might also be a sign of the Kitaev spin liquid.

Numerous theoretical studies have proposed effective Hamiltonians for $\alpha$-RuCl${}_3$, which have all revealed a Kitaev term \cite{winter2016,winter:2017breakdown, kim:2016crystal,kim:2015kitaev,wang:2017theoretical,hou2017}, and an important role is played by the symmetric off-diagonal exchange term $\Gamma$ \cite{winter2016,winter:2017breakdown}. As summarized in Ref.~\cite{winter:2017breakdown}, the differences in model proposals comes from the fact that there are many different parameters allowed by symmetry for an effective Hamiltonian, different crystal structures have been proposed (with $C2/m$ being preferred from more recent X-ray experiments \cite{cao:2016low,Kubota:2015Successive}), and the first-principle calculations depend heavily on interaction parameters. Much of the theoretical analysis of the above experiments has relied on spin wave theory (for below $T_N \approx 7$ K) and/or on calculations within the pure Kitaev model as it is accessible to quantum Monte Carlo simulations (see Ref.~\cite{nasu:2017thermal}, for example). 

Thermal conductivity measurements also provide a unique probe of this material and have generally seen ``unusual'' results \cite{Kasahara2017,kasahara:2018majorana,hentrich:2018large,leahy:2017anomalous}. A subset of the thermal conductivity data is an observation of a sizable thermal Hall conductivity $\kappa_{xy}$ \cite{Kasahara2017,kasahara:2018majorana,hentrich:2018large}, which,
 when compared with the theoretical predictions of the thermal hall conductivity of a pure Kitaev model at non-zero temperature \cite{nasu:2017thermal}, perhaps suggests the Kitaev-magnet nature of $\alpha$-RuCl${}_3$. Additionally, in the case of a large in-plane magnetic field, $\kappa_{xy}/T$ is reportedly quantized \cite{kasahara:2018majorana} at the same value that would be expected from the pure Kitaev model for a Majorana edge mode, $\pi k_B^2/(12 \hbar)$ \cite{kitaev:2006anyons}. 
 
 Though the data is not perfectly quantized, that may be explained due to interaction with phonons \cite{vinkler:2018approximately,ye:2018quantization}. Theoretical investigation has tried to explain the quantized value by finding the spin-liquid ground state of proposed Hamiltonians through a variational Monte Carlo method \cite{Liu:2018Dirac}. It was shown that a $\mathbb Z_2$ spin-liquid, as in the Kitaev model, is not preferred within the proposed $K-\Gamma$ model (see Eq.~\eqref{eq:Ham} and Table~\ref{params}), and the authors of Ref.~\cite{Liu:2018Dirac} claim that such a result will be true for a $J_1-K-\Gamma-J_3$ model too.

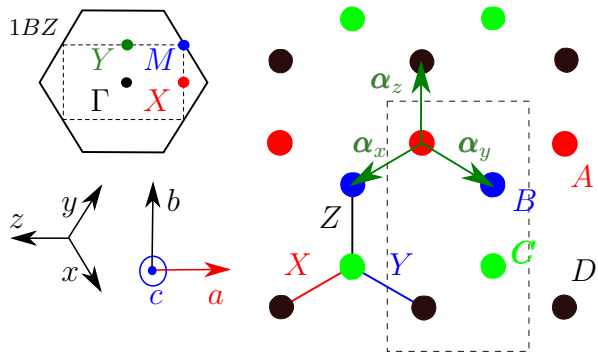
\begin{figure}
    \centering
    \begin{overpic}[width=.9\columnwidth]{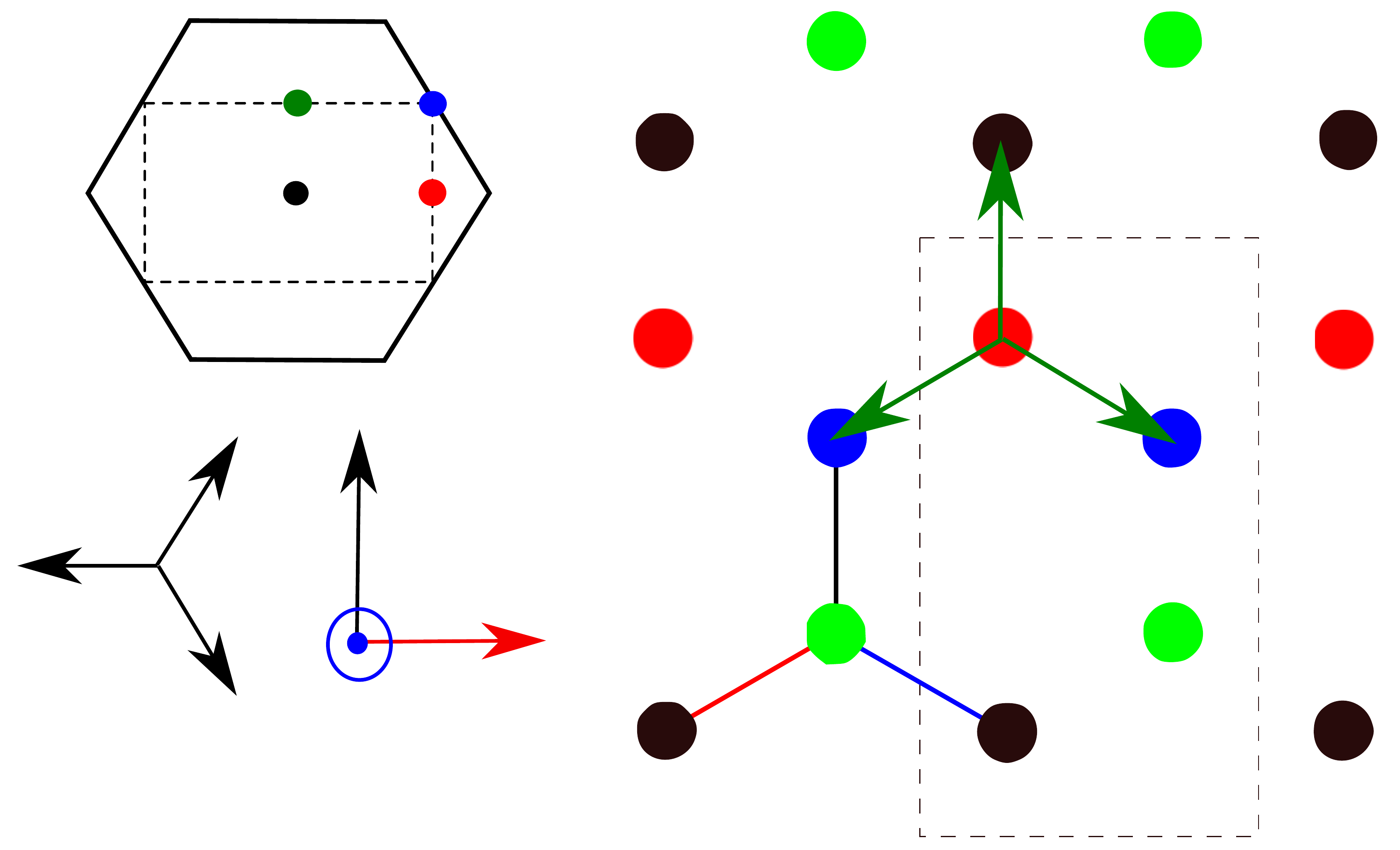}
    \put(54,22){\large $Z$}
    \put(48,14){\large \textcolor{red}{$X$}}
    \put(66,14){\large \textcolor{blue}{$Y$}}
    \put(97,13){\large $D$}
    \put(97,29){\large \textcolor{red}{$A$}}
    \put(87,25){\large \textcolor{blue}{$B$}}
    \put(87,16){\large \textcolor{green}{$\pmb C$}}
    \put(63,46){\large \textcolor{OliveGreen}{$\pmb \alpha_z$}}
    \put(60,35){\large \textcolor{OliveGreen}{$\pmb \alpha_x$}}
    \put(78,35){\large \textcolor{OliveGreen}{$\pmb \alpha_y$}}
    \put(1,22){\large $z$}
    \put(10,13){\large $x$}
    \put(10,25){\large $y$}
    \put(28,25){\large $b$}
    \put(35,9){\large \textcolor{red}{$a$}}
    \put(25,9){\large \textcolor{blue}{$c$}}
    \put(15,42){\large $\Gamma$}
    \put(24,42){\large \textcolor{red}{$X$}}
    \put(24,49){\large \textcolor{blue}{$M$}}
    \put(15,49){\large \textcolor{OliveGreen}{$Y$}}
    \put(1,55){$1BZ$}
    \end{overpic}
    \caption{(Color online). The conventions we use. The honeycomb lattice is shown on the right with the $X, Y,$ and $Z$ bond types labeled. The dashed line shows a unit cell containing one point in each of the four sub-lattices, $\{A,B,C,D\}$. The first Brillouin zone (1BZ) is shown in the upper left with four points labeled. The dashed line represents the 1BZ corresponding to the dashed box on the honeycomb lattice where the solid line represents the 1BZ of the triangular lattice. There are two coordinate systems $xyz$ and $abc$ that are shown in the bottom left. Note $c$ is in the [111] direction and $a$ is in the $[11\bar 2]$ direction. We fix the lattice spacing to be $1$ so that $|\alpha_i|=1$.}
    \label{fig:conventions}
\end{figure}
 
 Interestingly, relative to the quantized value, the data of Ref.~\cite{Kasahara2017} also suggests a large magnitude $\kappa_{xy}/T$ {\it even in the magnetically ordered state} of the material below $T_N\approx 7$ K, with sign reversed relative to the putative spin liquid state. Although the diagonal thermal conductivity $\kappa_{xx}$ receives a large contribution from phonons, the phonon thermal Hall conductivity at $T\lesssim 10$ K and $\mu_0 H \lesssim 15$ T has been measured for multiple other materials to be small ($\kappa_{xy} \lesssim 10^{-4}$ W/K/m) \cite{Inyushkin:2007phonon,Sugii:2017phonon}. It is well known that Berry curvature of the magnon bands can induce a finite thermal Hall effect \cite{matsumoto2014,lee2017}, and it is well studied that honeycomb Hamiltonians can have thermal Hall effects \cite{owerre:2016topological,owerre2016magnon}. A recent study has also looked at the thermal Hall conductivity of Kitaev materials in high-fields \cite{mcclarty:2018topological}.

 To our knowledge, however, no theoretical calculation of the thermal Hall conductivity due to magnons for low magnetic fields for $\alpha$-RuCl${}_3$ has been carried out. Taking the low-temperature data of Ref.~\cite{Kasahara2017} at face value, it provides a new test of any proposed effective Hamiltonian. In this work, we carry out such a calculation via linear spin wave theory (SWT). Though SWT may be unable to capture some features of the aforementioned experiments, it has been seen to well explain the THz spectroscopy data \cite{wu:2018field}. Indeed, in this work, we find that no previously proposed model predicts a large enough magnon thermal Hall effect, but, by decreasing the strength of a third nearest-neighbor Heisenberg interaction, such a model can be found.  The conclusion is that the Berry phase of magnon bands could explain the experimental observations in the ordered phase, but only if the effective Hamiltonian is somewhat different from previous proposals.

\section{Spin Wave Theory}
Many effective Hamiltonian models have been proposed for $\alpha$-RuCl${}_3$ (see Table~\ref{params}), and most of them can be captured by a $J_1-K-\Gamma-\Gamma'-J_3$ Hamiltonian:
\begin{equation}
\begin{aligned}
H&=\sum_{\langle i j\rangle} J_1 {\bf S_i} \cdot {\bf S_j} +K S_i^\gamma S_j^\gamma+\Gamma (S_i^\alpha S_j^\beta +S_i^\beta S_j^\alpha)
\\&+\Gamma'\left[S_i^\gamma (S_j^\alpha+S_j^\beta)+S_j^\gamma (S_i^\alpha+S_i^\beta)\right]
\\&+ \sum_{\langle\langle\langle i j \rangle \rangle \rangle} J_3 {\bf S}_i\cdot {\bf S}_j - \sum_i {\bf h} \cdot {\bf S}_i
\end{aligned}
\label{eq:Ham}
\end{equation}
where ${\bf h} = g \mu_B \mu_0 {\bf H}$ with $\mu_B$ the Bohr magneton and $g$ the $g$-factor. The index $\gamma(i,j)$ refers to the bond type of $(i,j)$ as indicated in Fig.~\ref{fig:conventions}, and $\alpha,\beta$ refer to the other two coordinates (e.g. $(\alpha,\beta)=(x,y)$ if $\gamma=z$). Note that the axes are arranged such that the, e.g., $X$ bond type is perpendicular to the $x$-axis. 

Proceeding with a standard spin wave theory (SWT) analysis following Ref.~\cite{Jones}, we can arrive at a Hamiltonian of free magnons (see Supplemental Material).
We allow the spin moments to be pointed in arbitrary directions along the four sublattices indicated in Fig.~\ref{fig:conventions}. We introduce four types of Holstein-Primakov bosons $b_i^{X}$ on the $X \in \{A,B,C,D\}$ sublattice at point $r_i$. We fix the angles by insisting on having no linear term in the boson creation/annihilation operators, and we always look for a zigzag solution (i.e. $\theta_A = \theta_B, \theta_C =\theta_D, \phi_A=\phi_B$ and $\phi_C =\phi_D$).  Defining $\psi_k^\dagger = (b_k^{A,\dagger},b_k^{B,\dagger},b_k^{C,\dagger},b_k^{D,\dagger},b_{-k}^A,b_{-k}^B,b_{-k}^C,b_{-k}^D)$, we can write the Hamiltonian, up to a constant, as
\begin{equation}
H = \frac12\sum_k \psi_k^\dagger \mathcal H \psi_k = \sum_k \sum_{n=1}^4 \omega_n \left(\gamma_{k,n}^\dagger \gamma_{k,n}+\frac12\right),
\end{equation}
where, in the last step, we perform a Bogoliubov transformation $(\vec \gamma_k,\vec \gamma_{-k}^\dagger)^T=\phi_k = T_k^{-1} \psi_k$ to diagonalize the Hamiltonian. To satisfy the boson commutation relations
\begin{equation}
    \sigma_3 T_k^{\dagger} \sigma_3 = T_{k}^{-1};\qquad \sigma_3 = \begin{pmatrix} 1_{4\times 4} & 0 \\ 0 & -1_{4 \times 4} \end{pmatrix}.
\end{equation}

 To ensure the validity of SWT, we compute the spin reduction for the $i$th boson
\begin{equation}
\Delta S_{0,i} = \frac{1}{V_\text{1BZ}}\int_{\text{1BZ}}d^2k \left( \sum_{j=1}^8 |T_{ij}|^2 n_\text{BE}(|\omega_j|) + \sum_{j=4}^8 |T_{ij}|^2\right),
\end{equation}
where $n_\text{BE}(\omega)=1/(e^{\omega/(k_BT)}-1)$ is the Bose-Einstein distribution, and $V_\text{1BZ}$ is the volume of the (2D) first Brillouin Zone (1BZ).
The reduction is often significant, but is always less than $70\%$, unless otherwise mentioned.

We can use SWT to find the magnon thermal Hall conductivity using an expression derived in Ref.~\cite{matsumoto2014} through linear response theory:

\begin{equation}\label{eq:kxy}
\begin{aligned}
\frac{ \kappa_{xy}}{k_B^2T/\hbar}&
=-\frac{1/d_c}{(2\pi)^2}\int_{\text{1BZ}} d^2k \sum_{n=1}^{4} \left(c_2(n_\text{BE}(\omega_n))-\frac{\pi^2}3\right) \Omega_{nn}\\
\Omega_{nn} &= \left(i \epsilon_{\mu \nu c}\sigma_3\frac{\partial T_k^\dagger}{\partial k_\mu}\sigma_3\frac{\partial T_k}{\partial k_\nu}\right)_{nn}
\end{aligned}
\end{equation}
where $\epsilon_{ijk}$ is the Levi-Civita symbol, $d_c$ is the distance between 2D planes (assuming well separated layers), $\Omega_{nn}$ is the Berry curvature of the $n$th magnon band, and, with $\text{Li}_2(z)$ being the dilogarithm,
\begin{equation}
c_2 (x) =(1+x) \left( \ln \frac{1+x}{x} \right)^2 - (\ln x)^2 -2 \text{Li}_2(-x) 
\end{equation}
\section{Results and Discussion}

We now specify the parameters in Eq.~\eqref{eq:Ham} to perform the calculation. We consider many different parameter sets as have been proposed in the literature, which are summarized in Table~\ref{params}. Though there is some experimental disagreement in the $g$-factor \cite{Kubota:2015Successive,yadav:2016kitaev}, susceptibility measurements \cite{trebst:2017kitaev,sears:2015moment,majumder:2015anisotropic} give the paramagnetic moment to be $\gtrsim 2\mu_B$, with $S=1/2$. We therefore fix $g=2.3$ to get the correct order of magnitude as in Refs.~\cite{winter2016,winter:2017breakdown,wu:2018field}, and $S=1/2$.  According to Refs.~\cite{Kasahara2017,johnson:2015monoclinic}, the interplanar distance is $d_c=5.72$ \AA.

\begin{table}
	\centering
\begin{tabular}{l| ccccccc|l}
	Name & $J_1$ & $K$ & $\Gamma$ & $\Gamma'$  & $J_2$ & $J_3$ & $K_3$ & Ref. \\
	\hline
HK	&   $-4.6$    &   $7.0$  & $-$ & $-$      & $-$ &   $-$   &    $-$ & \cite{banerjee:2016proximate}      \\
K$\Gamma$	&   $-$ &  $-6.8$   &  $9.5$   &  $-$&  $-$ & $-$  &   $-$   &    \cite{ran:2017spin}   \\
1(HK$\Gamma$J3)	&   $-1.7$    & $-6.7$    & $6.6$ & $-$ & $-$    & $2.7$   & $-$ & \cite{winter2016}          \\
2(HK$\Gamma$J3) & $-5.5$ & $7.6$ & $8.4$ & $-$ & $-$ &$2.3$ & $-$ & \cite{winter2016}\\
3(HK$\Gamma$J3) & $-0.5$ & $-5.0$ & $2.5$ & $-$ & $-$ & $0.5$ &  $-$ & \cite{winter:2017breakdown}\\
4(HK$\Gamma$J3)& $0.1$ & $-5.5$ & $7.6$ & $-$ &$-$ & $0.1$ & $-$  & \cite{wang:2017theoretical} \\
5(HK$\Gamma$J3) & $-0.3$  &$-10.9$ & $6.1$ & $-$ & $-$ & $0.03$ & $-$ & \cite{wang:2017theoretical}\\
6(HK$\Gamma$J3)	&    $-3.5$   &  $4.6$   &  $6.4$   &  $-$ & $-$ &  $0.8$ &   $-$    &    \cite{kim:2016crystal}     \\
(HK$\Gamma\Gamma$')	&    $-1$   &  $-8$   &  $4$   &  $-0.95$ & $-$ &  $-$ &   $-$    &    \cite{kim:2016crystal}     \\
HK$\Gamma$	&  $-12$     &  $17$   &  $12$   &  $-$  & $-$  &  $-$     &   $-$ &    \cite{kim:2015kitaev}    \\
HK$\Gamma$K3 &  $-1.8$ & $-10.6$ & $3.8$ & $-$ & $-$ & $1.25$ & $0.65$ & \cite{hou2017} \\
HK$\Gamma$J2 & $1.2$ & $-5.6$ & $1.0$ & $-$ & $0.3$  &$0.3$ & $-$ & \cite{yadav:2016kitaev} \\
\hline
7(HK$\Gamma$J3) & $-0.5$ & $-5.0$ & $2.5$ & $-$ & $-$ & $0.1125$ & $-$ & this paper
\end{tabular}
\caption{This information is primarily drawn from Table~1 in Ref.~\cite{paramsum} with some models added. All values are in meV. For this analysis, we ignore the $K_3,J_2$ values, which are there for completeness. Some groups propose different models within the same paper depending on the space group symmetry. For the HK and K$\Gamma$ model, we add a small $\Gamma$ and $J_3$ term, respectively to help the numerics.}
\label{params}
\end{table}

\begin{figure*}
    \centering
    \begin{overpic}[width=\columnwidth]{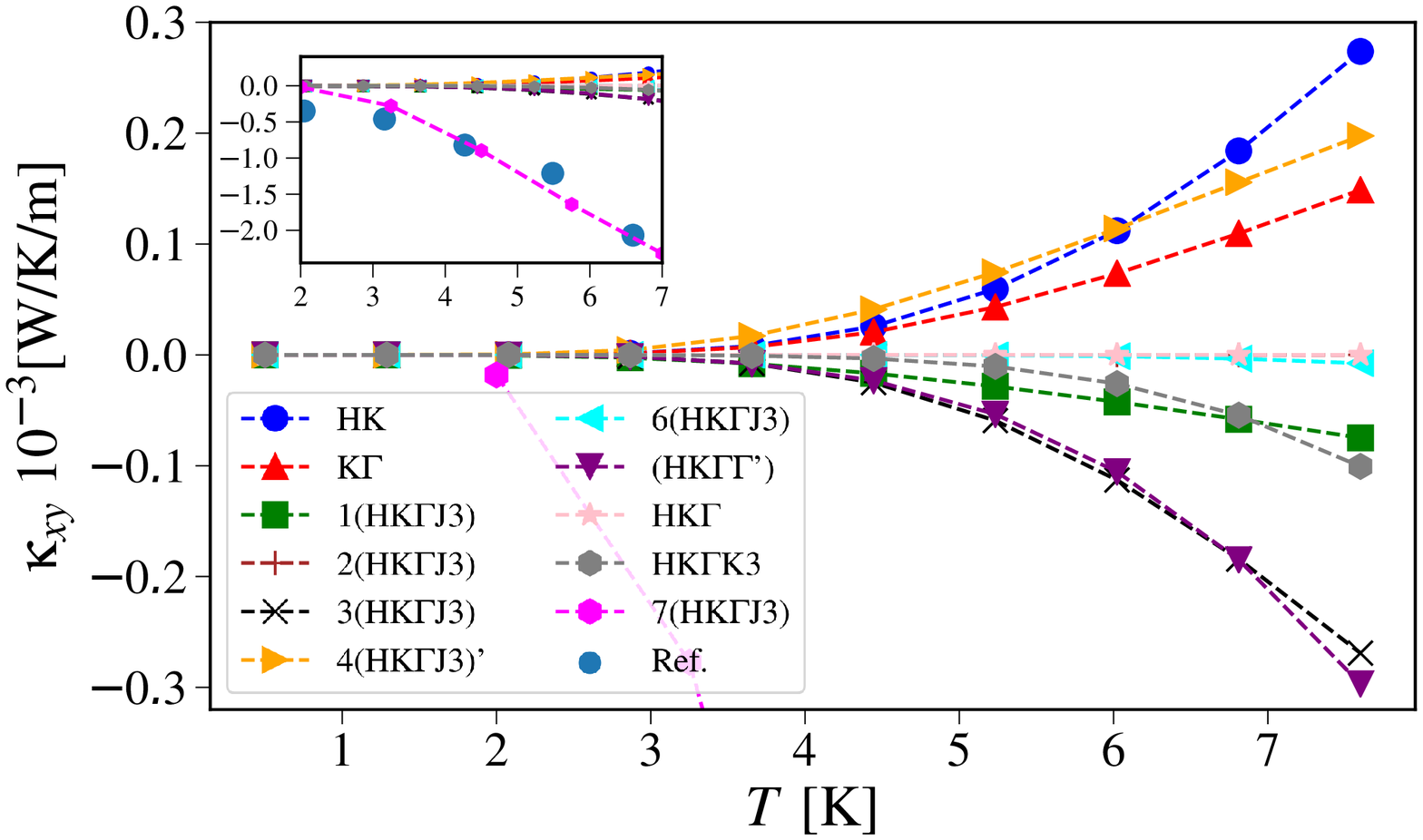}
    \put(60,14){\large $\mu_0 H$=12 [T]}
    \put(0,55){(a)}
    \put(50,14.6){\tiny \cite{Kasahara2017}}
    \end{overpic}
    \begin{overpic}[width=\columnwidth]{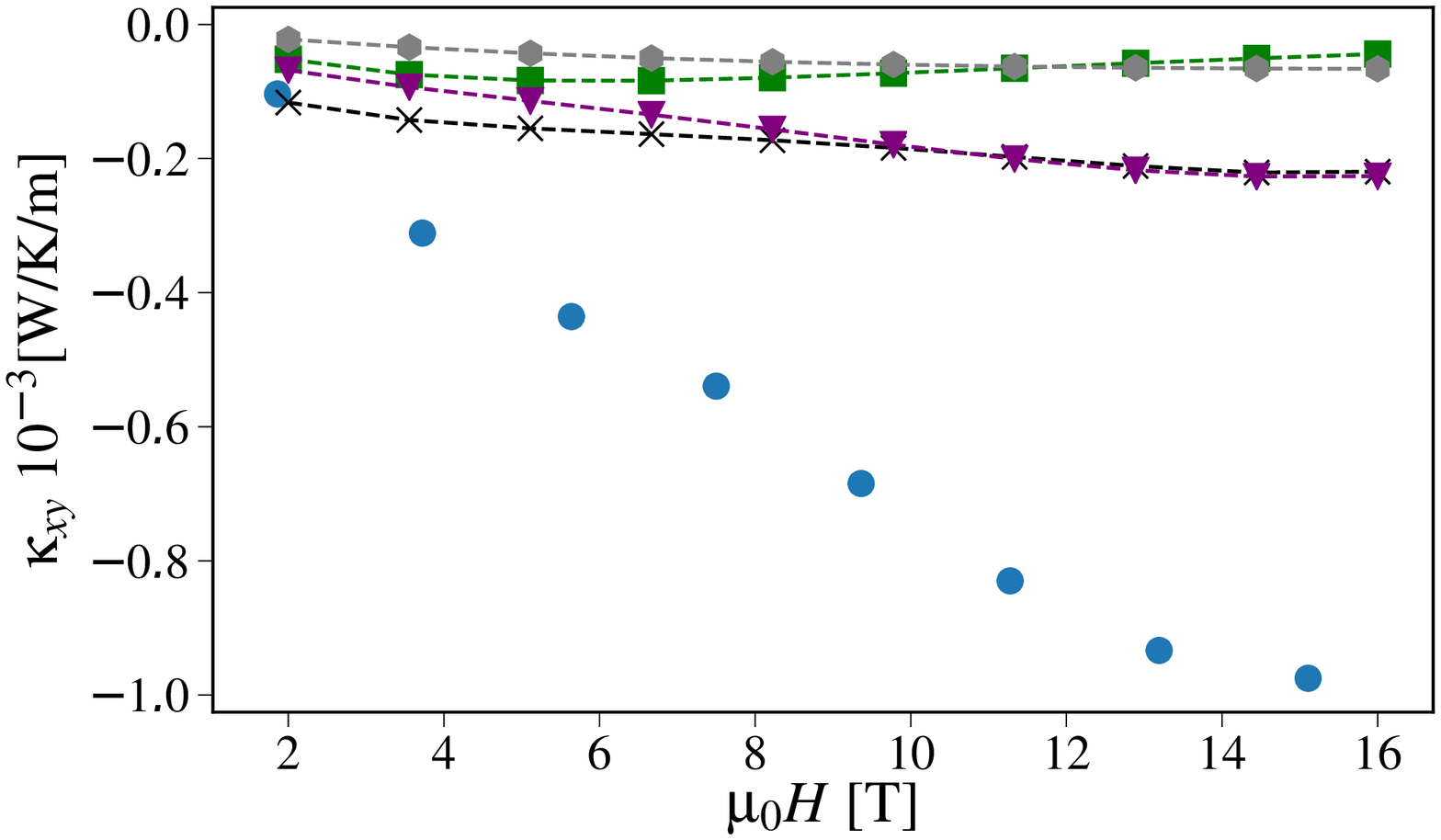}
    \put(18,14){\large $T = 7$ [K]}
    \put(0,55){(b)}
    \end{overpic}
    \caption{(Color online). We plot $\kappa_{xy}$ as computed from Eq.~\eqref{eq:kxy} for the various models in Table~\ref{params} as a function of (a) temperature and (b) magnetic field. We also plot the data from Ref.~\cite{Kasahara2017} as blue dots. The inset of (a) shows a zoomed-out version of the same graph. In (b), our model and models with $\kappa_{xy}\gtrsim0$ were removed. We do not plot 5(HK$\Gamma$J3) or HK$\Gamma$J2 since the zigzag spin wave solution becomes unstable for some critical magnetic field $\mu_0 H < 10$ T. Our proposed model, 7(HK$\Gamma$J3) has a large spin reduction $\Delta S_0/S \sim 0.9$ at $T=7$ K. }
    \label{fig:Kascompare}
\end{figure*}

The results of our SWT calculation are presented in Figs.~\ref{fig:Kascompare}. We have compared our code with the results of Ref.~\cite{mcclarty:2018topological} and Ref.~\cite{owerre:2016topological} to verify correctness. We have also plotted the data from Ref.~\cite{Kasahara2017}. We do not plot 5(HK$\Gamma$J3) or HK$\Gamma$J2 since the spin wave solution is not stable (i.e. there are complex eigenvalues) above some critical field $\mu_0 H<10$ T.

We see rather poor agreement between the models and the theory. Although most models do predict $\kappa_{xy}$ of the correct sign, models HK, K$\Gamma$, 4(HK$\Gamma$J3) do not. Further notice that all models with $K>0$ predict $\kappa_{xy} \gtrsim 0$.

To investigate why there is such a large discrepancy between the theoretical $\kappa_{xy}$ and the data of Ref.~\cite{Kasahara2017}, we try to find a large $\kappa_{xy}$ in a minimal $J_1-K-\Gamma-J_3$ model. It is worth noting that there is not much freedom. From Curie-Weiss temperature data $|K| \sim 100$ K $= 8.6 $ meV  \cite{trebst:2017kitaev,hirobe:2017magnetic}, which is similar to the estimate of Ref.~\cite{Kasahara2017} and is commonly seen in almost all of the models in Table~\ref{params}. Furthermore, it has been observed that the magnetic moments lie in the $ac$ plane and make an angle of approximately $35^\circ$ \cite{cao:2016low}, which requires a particular $\Gamma/K$. Minimizing the classical energy assuming the moments are in the $[xxz]$ direction,  in the $J_1-K-\Gamma-J_3$ model, we obtain an expression equivalent to one in Ref.~\cite{paramsum}:
\begin{equation}
    \frac{\Gamma}{K} = \frac{2}{\sqrt{2}\tan(\theta)+1-\sqrt{2}\cot(\theta)},
\end{equation}
where $z=\cos(\theta)$ and we assume $x>0$. Two minima of the classical energy can be found with $K<0; \Gamma/K \approx -0.82$ and $K>0; \Gamma/K \approx 0.0065$. 

Therefore, for two values of $\Gamma/K$ with differing signs of $K$, we have only freedom in $J_1$ and $J_3$. $J_1 <0$ and $J_3>0$ help stabilize the zigzag order, so we place these constraints. In the $K>0$ case with large enough $|J_1|$ to stabilize the zigzag order, we always found $\kappa_{xy}\gtrsim0$, though a more thorough search of the parameter space might be needed.

In the $K<0$ case, we instead start with the results of the meta-analysis of {\it ab initio} models from Ref.~\cite{winter:2017breakdown}: $\Gamma/|K| \approx 0.5$ and $J_1/|K| \approx 0.1$. Fixing $K= -5$ meV as in their proposed model, we scan possible values of $J_3$. We find that sufficiently low $J_3$ leads to large enough $\kappa_{xy}$ to explain all but the lowest temperature point of Ref.~\cite{Kasahara2017}. As a representative model, we find that $J_3 = 0.1125$ does well to reproduce the temperature data, as is shown in Fig.~\ref{fig:Kascompare}. Generically, with $-5$ meV $\gtrsim K \gtrsim -8$ meV, there is a value of $J_3 \gtrsim |K|/200$ that provides an order-of-magnitude fit to the data. For small $J_3$, though, we find a large spin reduction with $\Delta S_0/S \sim 0.9$. Further, these models predict much larger $\kappa_{xy}$ at $T=7$ K than is measured in Ref.~\cite{Kasahara2017}. Because of the proximity to the temperature at which long range order is lost, it is perhaps expected that whatever process is creating a large positive $\kappa_{xy}$ above $T_N \sim 7$ K is beginning to affect the conductivity at $T = 7$ K. We are making no claim that our model fits other experimental data.

\begin{figure}
    \centering
    \includegraphics[width=\columnwidth]{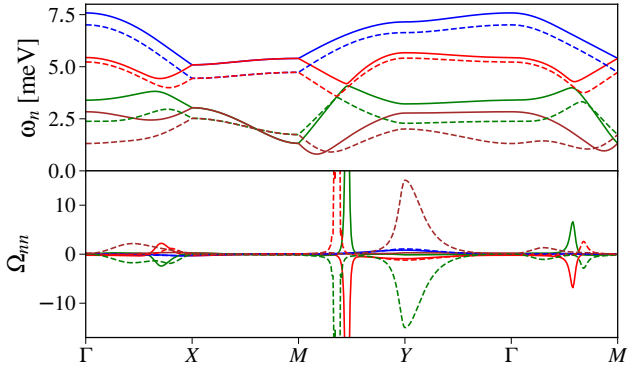}
    \caption{(Color online). See Fig.~\ref{fig:conventions} for the naming of the 1BZ points. We plot the SWT bands, $\omega_n$, and the Berry curvature, $\Omega_{nn}$, for the 3(HK$\Gamma$J3) model (solid lines) and the 7(HK$\Gamma$J3) model (dashed lines) at $\mu_0 H = 12$ T. The latter model has a much larger $\kappa_{xy}$ as seen in Fig.~\ref{fig:Kascompare} due to the fact that a) the gap between the lower two bands around the $Y$ point is smaller and at lower energy and b) the gap between the green and red band is at a lower energy. Note that the Berry curvature is largest where the band gap is smallest, and the lower energy means that the effect of $-c_2(n_\text{BE}(\omega_n))$ is more significant. (For $\Omega_{nn}$, the path is changed to be slightly inside the 1BZ as opposed to being on the boundary, when applicable.)}
    \label{fig:exbands}
\end{figure}

To see why the $\kappa_{xy}$ increased, we plot in Fig.~\ref{fig:exbands} an example of the linear SWT bands and Berry curvature for a particular path through the 1BZ for our model 7(HK$\Gamma$J3) vs. the similar model 3(HK$\Gamma$J3) on which it is based.

The difference in magnitude of $\kappa_{xy}$ can be understood as follows. The function $f(\omega_n)=-(c_2(n_\text{BE}(\omega_n))-\pi^2/3)$ scaling the Berry curvature in the $\kappa_{xy}$ integrand essentially serves as a high-pass filter with frequency $\omega_{Hi}=k_B T$. That is, if $\omega_n\ll k_B T$, $f(\omega_n)\sim T/\omega_n$ and if $\omega_n\gg k_BT, f(\omega_n)\sim \pi^2/3$. Since the sum of the Berry curvature integrated over the 1BZ is zero \cite{matsumoto2014}, then $k_BT\gtrsim \omega_n$  for $\kappa_{xy}$ to be significant. Furthermore, we can see from Fig.~\ref{fig:exbands} that the Berry curvature is largest when there is a small gap between the bands. To make the largest possible $\kappa_{xy}$, there must be small gaps in the bands at energies $\omega_n \lesssim k_B T$. 

These observations, however, call into question the validity of the low temperature data of Ref.~\cite{Kasahara2017}. Since the mass of the magnons has been estimated to be $\sim 2$ meV from inelastic neutron scattering \cite{ran:2017spin}, there should not be a large $\kappa_{xy}$ at temperature less than $T \sim 2$ meV$/(10 k_B)=2.3$ K, yet Ref.~\cite{Kasahara2017} reports a fairly large $\kappa_{xy}$ at $2.2$ K. 

Our analysis tends to favor $J_3$ smaller than has been proposed. As can be seen in plots in Refs.~\cite{winter2016,yadav:2016kitaev}, decreasing $J_3$ tends to move closer to a transition out of the zigzag order. Since smaller band gaps lead to larger Berry curvature, this result makes sense as SWT would predict a magnetic ordering phase transition when a gap in two bands close: assuming the energies of the bands have the form $\omega_\pm = a \pm \sqrt{b}$, a transition occurs at $b=0$ since $b<0$ leads to instability of the spin wave solution. 

Another interpretation of our results is that the models in Table~\ref{params} are consistent with the data if the magnons are not the dominant source of the $\kappa_{xy}$ at low temperatures. Phonons could in principle  give a larger contribution than ordinarily observed. To elaborate, in past experiments, the Hall angle was measured at $\mu_0 H \sim 10$ T to be $\kappa_{xx}/\kappa_{xy} \sim 1-5 \times 10^{-4}$ \cite{Sugii:2017phonon,Inyushkin:2007phonon} and $\kappa_{xx}\sim 2-6$ W/m/K for $\alpha-$RuCl${}_3$ at $T\lesssim 10$ K \cite{Kasahara2017,leahy:2017anomalous,hentrich:2018large}. We would then estimate $\kappa_{xy}\sim 3 \times 10^{-4}\kappa_{xx} \approx 1.2 \times 10^{-3}$ W/m/K, which is of the correct order. Regarding contributions in the ordered phase by Kitaev-like excitations from fluctuations, note that since the pure Kitaev model predicts $\kappa_{xy}>0$ \cite{nasu:2017thermal}, and the experimental data shows that $\kappa_{xy}$ switches sign at $T\sim T_N$~\cite{Kasahara2017}, these observations would be inconsistent with having fluctuations into the Kitaev model explain the discrepancy between the predicted and observed low temperature $\kappa_{xy}$.

\section{Conclusions}

We have investigated whether the thermal Hall conductivity data reported in Ref.~\cite{Kasahara2017} at low temperatures for the material $\alpha$-RuCl${}_3$ can be explained through linear SWT as an example of the magnon thermal Hall effect. Although for many of the effective Hamiltonians proposed in the literature, we find a non-zero $\kappa_{xy}$ of the correct sign, none could satisfactorily explain the data. By modifying the model of Ref.~\cite{winter:2017breakdown}, we were able to find large enough $\kappa_{xy}$ to explain the $7$ K $\gtrsim T\gtrsim 3$ K data showing that it is possible to explain the data via linear SWT. It also appears that $K>0$ is not favored solely based on the sign of $\kappa_{xy}$. Taking the data at face value, these measurements provide a novel way to constrain a proposed effective Hamiltonian. In the future, more experimental data, better theoretical methods to incorporate both spin waves and Kitaev-model quasiparticles, and more careful treatment of the phonon thermal Hall effect would be useful in constraining proposed effective Hamiltonians.

{\it Acknowledgements:} The authors thank  J. W. Orenstein for useful conversations.  We acknowledge support from the Quantum Materials Program at LBNL, funded by the US Department of Energy under Contract No. DE-AC02-05CH11231. 

\bibliography{SpinWaveRuCl3Bib.bib}

\begin{thebibliography}{44}%
\makeatletter
\providecommand \@ifxundefined [1]{%
 \@ifx{#1\undefined}
}%
\providecommand \@ifnum [1]{%
 \ifnum #1\expandafter \@firstoftwo
 \else \expandafter \@secondoftwo
 \fi
}%
\providecommand \@ifx [1]{%
 \ifx #1\expandafter \@firstoftwo
 \else \expandafter \@secondoftwo
 \fi
}%
\providecommand \natexlab [1]{#1}%
\providecommand \enquote  [1]{``#1''}%
\providecommand \bibnamefont  [1]{#1}%
\providecommand \bibfnamefont [1]{#1}%
\providecommand \citenamefont [1]{#1}%
\providecommand \href@noop [0]{\@secondoftwo}%
\providecommand \href [0]{\begingroup \@sanitize@url \@href}%
\providecommand \@href[1]{\@@startlink{#1}\@@href}%
\providecommand \@@href[1]{\endgroup#1\@@endlink}%
\providecommand \@sanitize@url [0]{\catcode `\\12\catcode `\$12\catcode
  `\&12\catcode `\#12\catcode `\^12\catcode `\_12\catcode `\%12\relax}%
\providecommand \@@startlink[1]{}%
\providecommand \@@endlink[0]{}%
\providecommand \url  [0]{\begingroup\@sanitize@url \@url }%
\providecommand \@url [1]{\endgroup\@href {#1}{\urlprefix }}%
\providecommand \urlprefix  [0]{URL }%
\providecommand \Eprint [0]{\href }%
\providecommand \doibase [0]{http://dx.doi.org/}%
\providecommand \selectlanguage [0]{\@gobble}%
\providecommand \bibinfo  [0]{\@secondoftwo}%
\providecommand \bibfield  [0]{\@secondoftwo}%
\providecommand \translation [1]{[#1]}%
\providecommand \BibitemOpen [0]{}%
\providecommand \bibitemStop [0]{}%
\providecommand \bibitemNoStop [0]{.\EOS\space}%
\providecommand \EOS [0]{\spacefactor3000\relax}%
\providecommand \BibitemShut  [1]{\csname bibitem#1\endcsname}%
\let\auto@bib@innerbib\@empty
\bibitem [{\citenamefont {Kitaev}(2006)}]{kitaev:2006anyons}%
  \BibitemOpen
  \bibfield  {author} {\bibinfo {author} {\bibfnamefont {A.}~\bibnamefont
  {Kitaev}},\ }\href {\doibase https://doi.org/10.1016/j.aop.2005.10.005}
  {\bibfield  {journal} {\bibinfo  {journal} {Annals of Physics}\ }\textbf
  {\bibinfo {volume} {321}},\ \bibinfo {pages} {2 } (\bibinfo {year} {2006})},\
  \bibinfo {note} {january Special Issue}\BibitemShut {NoStop}%
\bibitem [{\citenamefont {Jackeli}\ and\ \citenamefont
  {Khaliullin}(2009)}]{jackeli:2009mott}%
  \BibitemOpen
  \bibfield  {author} {\bibinfo {author} {\bibfnamefont {G.}~\bibnamefont
  {Jackeli}}\ and\ \bibinfo {author} {\bibfnamefont {G.}~\bibnamefont
  {Khaliullin}},\ }\href {\doibase 10.1103/PhysRevLett.102.017205} {\bibfield
  {journal} {\bibinfo  {journal} {Phys. Rev. Lett.}\ }\textbf {\bibinfo
  {volume} {102}},\ \bibinfo {pages} {017205} (\bibinfo {year}
  {2009})}\BibitemShut {NoStop}%
\bibitem [{\citenamefont {Trebst}(2017)}]{trebst:2017kitaev}%
  \BibitemOpen
  \bibfield  {author} {\bibinfo {author} {\bibfnamefont {S.}~\bibnamefont
  {Trebst}},\ }\href@noop {} {\bibfield  {journal} {\bibinfo  {journal} {arXiv
  preprint arXiv:1701.07056}\ } (\bibinfo {year} {2017})}\BibitemShut {NoStop}%
\bibitem [{\citenamefont {Koitzsch}\ \emph {et~al.}(2016)\citenamefont
  {Koitzsch}, \citenamefont {Habenicht}, \citenamefont {M\"uller},
  \citenamefont {Knupfer}, \citenamefont {B\"uchner}, \citenamefont {Kandpal},
  \citenamefont {van~den Brink}, \citenamefont {Nowak}, \citenamefont
  {Isaeva},\ and\ \citenamefont {Doert}}]{Koitzsch:2016jeff}%
  \BibitemOpen
  \bibfield  {author} {\bibinfo {author} {\bibfnamefont {A.}~\bibnamefont
  {Koitzsch}}, \bibinfo {author} {\bibfnamefont {C.}~\bibnamefont {Habenicht}},
  \bibinfo {author} {\bibfnamefont {E.}~\bibnamefont {M\"uller}}, \bibinfo
  {author} {\bibfnamefont {M.}~\bibnamefont {Knupfer}}, \bibinfo {author}
  {\bibfnamefont {B.}~\bibnamefont {B\"uchner}}, \bibinfo {author}
  {\bibfnamefont {H.~C.}\ \bibnamefont {Kandpal}}, \bibinfo {author}
  {\bibfnamefont {J.}~\bibnamefont {van~den Brink}}, \bibinfo {author}
  {\bibfnamefont {D.}~\bibnamefont {Nowak}}, \bibinfo {author} {\bibfnamefont
  {A.}~\bibnamefont {Isaeva}}, \ and\ \bibinfo {author} {\bibfnamefont
  {T.}~\bibnamefont {Doert}},\ }\href {\doibase 10.1103/PhysRevLett.117.126403}
  {\bibfield  {journal} {\bibinfo  {journal} {Phys. Rev. Lett.}\ }\textbf
  {\bibinfo {volume} {117}},\ \bibinfo {pages} {126403} (\bibinfo {year}
  {2016})}\BibitemShut {NoStop}%
\bibitem [{\citenamefont {Baek}\ \emph {et~al.}(2017)\citenamefont {Baek},
  \citenamefont {Do}, \citenamefont {Choi}, \citenamefont {Kwon}, \citenamefont
  {Wolter}, \citenamefont {Nishimoto}, \citenamefont {van~den Brink},\ and\
  \citenamefont {B\"uchner}}]{baek:2017evidence}%
  \BibitemOpen
  \bibfield  {author} {\bibinfo {author} {\bibfnamefont {S.-H.}\ \bibnamefont
  {Baek}}, \bibinfo {author} {\bibfnamefont {S.-H.}\ \bibnamefont {Do}},
  \bibinfo {author} {\bibfnamefont {K.-Y.}\ \bibnamefont {Choi}}, \bibinfo
  {author} {\bibfnamefont {Y.~S.}\ \bibnamefont {Kwon}}, \bibinfo {author}
  {\bibfnamefont {A.~U.~B.}\ \bibnamefont {Wolter}}, \bibinfo {author}
  {\bibfnamefont {S.}~\bibnamefont {Nishimoto}}, \bibinfo {author}
  {\bibfnamefont {J.}~\bibnamefont {van~den Brink}}, \ and\ \bibinfo {author}
  {\bibfnamefont {B.}~\bibnamefont {B\"uchner}},\ }\href {\doibase
  10.1103/PhysRevLett.119.037201} {\bibfield  {journal} {\bibinfo  {journal}
  {Phys. Rev. Lett.}\ }\textbf {\bibinfo {volume} {119}},\ \bibinfo {pages}
  {037201} (\bibinfo {year} {2017})}\BibitemShut {NoStop}%
\bibitem [{\citenamefont {Banerjee}\ \emph {et~al.}(2017)\citenamefont
  {Banerjee}, \citenamefont {Yan}, \citenamefont {Knolle}, \citenamefont
  {Bridges}, \citenamefont {Stone}, \citenamefont {Lumsden}, \citenamefont
  {Mandrus}, \citenamefont {Tennant}, \citenamefont {Moessner},\ and\
  \citenamefont {Nagler}}]{banerjee:2017neutron}%
  \BibitemOpen
  \bibfield  {author} {\bibinfo {author} {\bibfnamefont {A.}~\bibnamefont
  {Banerjee}}, \bibinfo {author} {\bibfnamefont {J.}~\bibnamefont {Yan}},
  \bibinfo {author} {\bibfnamefont {J.}~\bibnamefont {Knolle}}, \bibinfo
  {author} {\bibfnamefont {C.~A.}\ \bibnamefont {Bridges}}, \bibinfo {author}
  {\bibfnamefont {M.~B.}\ \bibnamefont {Stone}}, \bibinfo {author}
  {\bibfnamefont {M.~D.}\ \bibnamefont {Lumsden}}, \bibinfo {author}
  {\bibfnamefont {D.~G.}\ \bibnamefont {Mandrus}}, \bibinfo {author}
  {\bibfnamefont {D.~A.}\ \bibnamefont {Tennant}}, \bibinfo {author}
  {\bibfnamefont {R.}~\bibnamefont {Moessner}}, \ and\ \bibinfo {author}
  {\bibfnamefont {S.~E.}\ \bibnamefont {Nagler}},\ }\href {\doibase
  10.1126/science.aah6015} {\bibfield  {journal} {\bibinfo  {journal}
  {Science}\ }\textbf {\bibinfo {volume} {356}},\ \bibinfo {pages} {1055}
  (\bibinfo {year} {2017})}\BibitemShut {NoStop}%
\bibitem [{\citenamefont {Ran}\ \emph {et~al.}(2017)\citenamefont {Ran},
  \citenamefont {Wang}, \citenamefont {Wang}, \citenamefont {Dong},
  \citenamefont {Ren}, \citenamefont {Bao}, \citenamefont {Li}, \citenamefont
  {Ma}, \citenamefont {Gan}, \citenamefont {Zhang}, \citenamefont {Park},
  \citenamefont {Deng}, \citenamefont {Danilkin}, \citenamefont {Yu},
  \citenamefont {Li},\ and\ \citenamefont {Wen}}]{ran:2017spin}%
  \BibitemOpen
  \bibfield  {author} {\bibinfo {author} {\bibfnamefont {K.}~\bibnamefont
  {Ran}}, \bibinfo {author} {\bibfnamefont {J.}~\bibnamefont {Wang}}, \bibinfo
  {author} {\bibfnamefont {W.}~\bibnamefont {Wang}}, \bibinfo {author}
  {\bibfnamefont {Z.-Y.}\ \bibnamefont {Dong}}, \bibinfo {author}
  {\bibfnamefont {X.}~\bibnamefont {Ren}}, \bibinfo {author} {\bibfnamefont
  {S.}~\bibnamefont {Bao}}, \bibinfo {author} {\bibfnamefont {S.}~\bibnamefont
  {Li}}, \bibinfo {author} {\bibfnamefont {Z.}~\bibnamefont {Ma}}, \bibinfo
  {author} {\bibfnamefont {Y.}~\bibnamefont {Gan}}, \bibinfo {author}
  {\bibfnamefont {Y.}~\bibnamefont {Zhang}}, \bibinfo {author} {\bibfnamefont
  {J.~T.}\ \bibnamefont {Park}}, \bibinfo {author} {\bibfnamefont
  {G.}~\bibnamefont {Deng}}, \bibinfo {author} {\bibfnamefont {S.}~\bibnamefont
  {Danilkin}}, \bibinfo {author} {\bibfnamefont {S.-L.}\ \bibnamefont {Yu}},
  \bibinfo {author} {\bibfnamefont {J.-X.}\ \bibnamefont {Li}}, \ and\ \bibinfo
  {author} {\bibfnamefont {J.}~\bibnamefont {Wen}},\ }\href {\doibase
  10.1103/PhysRevLett.118.107203} {\bibfield  {journal} {\bibinfo  {journal}
  {Phys. Rev. Lett.}\ }\textbf {\bibinfo {volume} {118}},\ \bibinfo {pages}
  {107203} (\bibinfo {year} {2017})}\BibitemShut {NoStop}%
\bibitem [{\citenamefont {Leahy}\ \emph {et~al.}(2017)\citenamefont {Leahy},
  \citenamefont {Pocs}, \citenamefont {Siegfried}, \citenamefont {Graf},
  \citenamefont {Do}, \citenamefont {Choi}, \citenamefont {Normand},\ and\
  \citenamefont {Lee}}]{leahy:2017anomalous}%
  \BibitemOpen
  \bibfield  {author} {\bibinfo {author} {\bibfnamefont {I.~A.}\ \bibnamefont
  {Leahy}}, \bibinfo {author} {\bibfnamefont {C.~A.}\ \bibnamefont {Pocs}},
  \bibinfo {author} {\bibfnamefont {P.~E.}\ \bibnamefont {Siegfried}}, \bibinfo
  {author} {\bibfnamefont {D.}~\bibnamefont {Graf}}, \bibinfo {author}
  {\bibfnamefont {S.-H.}\ \bibnamefont {Do}}, \bibinfo {author} {\bibfnamefont
  {K.-Y.}\ \bibnamefont {Choi}}, \bibinfo {author} {\bibfnamefont
  {B.}~\bibnamefont {Normand}}, \ and\ \bibinfo {author} {\bibfnamefont
  {M.}~\bibnamefont {Lee}},\ }\href {\doibase 10.1103/PhysRevLett.118.187203}
  {\bibfield  {journal} {\bibinfo  {journal} {Phys. Rev. Lett.}\ }\textbf
  {\bibinfo {volume} {118}},\ \bibinfo {pages} {187203} (\bibinfo {year}
  {2017})}\BibitemShut {NoStop}%
\bibitem [{\citenamefont {Kasahara}\ \emph
  {et~al.}(2018{\natexlab{a}})\citenamefont {Kasahara}, \citenamefont
  {Ohnishi}, \citenamefont {Mizukami}, \citenamefont {Tanaka}, \citenamefont
  {Ma}, \citenamefont {Sugii}, \citenamefont {Kurita}, \citenamefont {Tanaka},
  \citenamefont {Nasu}, \citenamefont {Motome} \emph
  {et~al.}}]{kasahara:2018majorana}%
  \BibitemOpen
  \bibfield  {author} {\bibinfo {author} {\bibfnamefont {Y.}~\bibnamefont
  {Kasahara}}, \bibinfo {author} {\bibfnamefont {T.}~\bibnamefont {Ohnishi}},
  \bibinfo {author} {\bibfnamefont {Y.}~\bibnamefont {Mizukami}}, \bibinfo
  {author} {\bibfnamefont {O.}~\bibnamefont {Tanaka}}, \bibinfo {author}
  {\bibfnamefont {S.}~\bibnamefont {Ma}}, \bibinfo {author} {\bibfnamefont
  {K.}~\bibnamefont {Sugii}}, \bibinfo {author} {\bibfnamefont
  {N.}~\bibnamefont {Kurita}}, \bibinfo {author} {\bibfnamefont
  {H.}~\bibnamefont {Tanaka}}, \bibinfo {author} {\bibfnamefont
  {J.}~\bibnamefont {Nasu}}, \bibinfo {author} {\bibfnamefont {Y.}~\bibnamefont
  {Motome}},  \emph {et~al.},\ }\href@noop {} {\bibfield  {journal} {\bibinfo
  {journal} {arXiv preprint arXiv:1805.05022}\ } (\bibinfo {year}
  {2018}{\natexlab{a}})}\BibitemShut {NoStop}%
\bibitem [{\citenamefont {Sandilands}\ \emph {et~al.}(2015)\citenamefont
  {Sandilands}, \citenamefont {Tian}, \citenamefont {Plumb}, \citenamefont
  {Kim},\ and\ \citenamefont {Burch}}]{Sandilands:2015Scattering}%
  \BibitemOpen
  \bibfield  {author} {\bibinfo {author} {\bibfnamefont {L.~J.}\ \bibnamefont
  {Sandilands}}, \bibinfo {author} {\bibfnamefont {Y.}~\bibnamefont {Tian}},
  \bibinfo {author} {\bibfnamefont {K.~W.}\ \bibnamefont {Plumb}}, \bibinfo
  {author} {\bibfnamefont {Y.-J.}\ \bibnamefont {Kim}}, \ and\ \bibinfo
  {author} {\bibfnamefont {K.~S.}\ \bibnamefont {Burch}},\ }\href {\doibase
  10.1103/PhysRevLett.114.147201} {\bibfield  {journal} {\bibinfo  {journal}
  {Phys. Rev. Lett.}\ }\textbf {\bibinfo {volume} {114}},\ \bibinfo {pages}
  {147201} (\bibinfo {year} {2015})}\BibitemShut {NoStop}%
\bibitem [{\citenamefont {Nasu}\ \emph {et~al.}(2016)\citenamefont {Nasu},
  \citenamefont {Knolle}, \citenamefont {Kovrizhin}, \citenamefont {Motome},\
  and\ \citenamefont {Moessner}}]{nasu:2016fermionic}%
  \BibitemOpen
  \bibfield  {author} {\bibinfo {author} {\bibfnamefont {J.}~\bibnamefont
  {Nasu}}, \bibinfo {author} {\bibfnamefont {J.}~\bibnamefont {Knolle}},
  \bibinfo {author} {\bibfnamefont {D.~L.}\ \bibnamefont {Kovrizhin}}, \bibinfo
  {author} {\bibfnamefont {Y.}~\bibnamefont {Motome}}, \ and\ \bibinfo {author}
  {\bibfnamefont {R.}~\bibnamefont {Moessner}},\ }\href@noop {} {\bibfield
  {journal} {\bibinfo  {journal} {Nature Physics}\ }\textbf {\bibinfo {volume}
  {12}},\ \bibinfo {pages} {912} (\bibinfo {year} {2016})}\BibitemShut
  {NoStop}%
\bibitem [{\citenamefont {Banerjee}\ \emph {et~al.}(2016)\citenamefont
  {Banerjee}, \citenamefont {Bridges}, \citenamefont {Yan}, \citenamefont
  {Aczel}, \citenamefont {Li}, \citenamefont {Stone}, \citenamefont {Granroth},
  \citenamefont {Lumsden}, \citenamefont {Yiu}, \citenamefont {Knolle} \emph
  {et~al.}}]{banerjee:2016proximate}%
  \BibitemOpen
  \bibfield  {author} {\bibinfo {author} {\bibfnamefont {A.}~\bibnamefont
  {Banerjee}}, \bibinfo {author} {\bibfnamefont {C.}~\bibnamefont {Bridges}},
  \bibinfo {author} {\bibfnamefont {J.-Q.}\ \bibnamefont {Yan}}, \bibinfo
  {author} {\bibfnamefont {A.}~\bibnamefont {Aczel}}, \bibinfo {author}
  {\bibfnamefont {L.}~\bibnamefont {Li}}, \bibinfo {author} {\bibfnamefont
  {M.}~\bibnamefont {Stone}}, \bibinfo {author} {\bibfnamefont
  {G.}~\bibnamefont {Granroth}}, \bibinfo {author} {\bibfnamefont
  {M.}~\bibnamefont {Lumsden}}, \bibinfo {author} {\bibfnamefont
  {Y.}~\bibnamefont {Yiu}}, \bibinfo {author} {\bibfnamefont {J.}~\bibnamefont
  {Knolle}},  \emph {et~al.},\ }\href@noop {} {\bibfield  {journal} {\bibinfo
  {journal} {Nature materials}\ }\textbf {\bibinfo {volume} {15}},\ \bibinfo
  {pages} {733} (\bibinfo {year} {2016})}\BibitemShut {NoStop}%
\bibitem [{\citenamefont {Little}\ \emph {et~al.}(2017)\citenamefont {Little},
  \citenamefont {Wu}, \citenamefont {Lampen-Kelley}, \citenamefont {Banerjee},
  \citenamefont {Patankar}, \citenamefont {Rees}, \citenamefont {Bridges},
  \citenamefont {Yan}, \citenamefont {Mandrus}, \citenamefont {Nagler},\ and\
  \citenamefont {Orenstein}}]{Little:2017Antiferromagnetic}%
  \BibitemOpen
  \bibfield  {author} {\bibinfo {author} {\bibfnamefont {A.}~\bibnamefont
  {Little}}, \bibinfo {author} {\bibfnamefont {L.}~\bibnamefont {Wu}}, \bibinfo
  {author} {\bibfnamefont {P.}~\bibnamefont {Lampen-Kelley}}, \bibinfo {author}
  {\bibfnamefont {A.}~\bibnamefont {Banerjee}}, \bibinfo {author}
  {\bibfnamefont {S.}~\bibnamefont {Patankar}}, \bibinfo {author}
  {\bibfnamefont {D.}~\bibnamefont {Rees}}, \bibinfo {author} {\bibfnamefont
  {C.~A.}\ \bibnamefont {Bridges}}, \bibinfo {author} {\bibfnamefont {J.-Q.}\
  \bibnamefont {Yan}}, \bibinfo {author} {\bibfnamefont {D.}~\bibnamefont
  {Mandrus}}, \bibinfo {author} {\bibfnamefont {S.~E.}\ \bibnamefont {Nagler}},
  \ and\ \bibinfo {author} {\bibfnamefont {J.}~\bibnamefont {Orenstein}},\
  }\href {\doibase 10.1103/PhysRevLett.119.227201} {\bibfield  {journal}
  {\bibinfo  {journal} {Phys. Rev. Lett.}\ }\textbf {\bibinfo {volume} {119}},\
  \bibinfo {pages} {227201} (\bibinfo {year} {2017})}\BibitemShut {NoStop}%
\bibitem [{\citenamefont {Wang}\ \emph
  {et~al.}(2017{\natexlab{a}})\citenamefont {Wang}, \citenamefont {Reschke},
  \citenamefont {H\"uvonen}, \citenamefont {Do}, \citenamefont {Choi},
  \citenamefont {Gensch}, \citenamefont {Nagel}, \citenamefont {R\~o\ om},\
  and\ \citenamefont {Loidl}}]{Wang:2017Magnetic}%
  \BibitemOpen
  \bibfield  {author} {\bibinfo {author} {\bibfnamefont {Z.}~\bibnamefont
  {Wang}}, \bibinfo {author} {\bibfnamefont {S.}~\bibnamefont {Reschke}},
  \bibinfo {author} {\bibfnamefont {D.}~\bibnamefont {H\"uvonen}}, \bibinfo
  {author} {\bibfnamefont {S.-H.}\ \bibnamefont {Do}}, \bibinfo {author}
  {\bibfnamefont {K.-Y.}\ \bibnamefont {Choi}}, \bibinfo {author}
  {\bibfnamefont {M.}~\bibnamefont {Gensch}}, \bibinfo {author} {\bibfnamefont
  {U.}~\bibnamefont {Nagel}}, \bibinfo {author} {\bibfnamefont
  {T.}~\bibnamefont {R\~o\ om}}, \ and\ \bibinfo {author} {\bibfnamefont
  {A.}~\bibnamefont {Loidl}},\ }\href {\doibase 10.1103/PhysRevLett.119.227202}
  {\bibfield  {journal} {\bibinfo  {journal} {Phys. Rev. Lett.}\ }\textbf
  {\bibinfo {volume} {119}},\ \bibinfo {pages} {227202} (\bibinfo {year}
  {2017}{\natexlab{a}})}\BibitemShut {NoStop}%
\bibitem [{\citenamefont {Wu}\ \emph {et~al.}(2018)\citenamefont {Wu},
  \citenamefont {Little}, \citenamefont {Aldape}, \citenamefont {Rees},
  \citenamefont {Thewalt}, \citenamefont {Lampen-Kelley}, \citenamefont
  {Banerjee}, \citenamefont {Bridges}, \citenamefont {Yan}, \citenamefont
  {Patankar} \emph {et~al.}}]{wu:2018field}%
  \BibitemOpen
  \bibfield  {author} {\bibinfo {author} {\bibfnamefont {L.}~\bibnamefont
  {Wu}}, \bibinfo {author} {\bibfnamefont {A.}~\bibnamefont {Little}}, \bibinfo
  {author} {\bibfnamefont {E.~E.}\ \bibnamefont {Aldape}}, \bibinfo {author}
  {\bibfnamefont {D.}~\bibnamefont {Rees}}, \bibinfo {author} {\bibfnamefont
  {E.}~\bibnamefont {Thewalt}}, \bibinfo {author} {\bibfnamefont
  {P.}~\bibnamefont {Lampen-Kelley}}, \bibinfo {author} {\bibfnamefont
  {A.}~\bibnamefont {Banerjee}}, \bibinfo {author} {\bibfnamefont {C.~A.}\
  \bibnamefont {Bridges}}, \bibinfo {author} {\bibfnamefont {J.}~\bibnamefont
  {Yan}}, \bibinfo {author} {\bibfnamefont {S.}~\bibnamefont {Patankar}},
  \emph {et~al.},\ }\href@noop {} {\bibfield  {journal} {\bibinfo  {journal}
  {arXiv preprint arXiv:1806.00855}\ } (\bibinfo {year} {2018})}\BibitemShut
  {NoStop}%
\bibitem [{\citenamefont {Ponomaryov}\ \emph {et~al.}(2017)\citenamefont
  {Ponomaryov}, \citenamefont {Schulze}, \citenamefont {Wosnitza},
  \citenamefont {Lampen-Kelley}, \citenamefont {Banerjee}, \citenamefont {Yan},
  \citenamefont {Bridges}, \citenamefont {Mandrus}, \citenamefont {Nagler},
  \citenamefont {Kolezhuk},\ and\ \citenamefont
  {Zvyagin}}]{Ponomaryov:2017Unconvetional}%
  \BibitemOpen
  \bibfield  {author} {\bibinfo {author} {\bibfnamefont {A.~N.}\ \bibnamefont
  {Ponomaryov}}, \bibinfo {author} {\bibfnamefont {E.}~\bibnamefont {Schulze}},
  \bibinfo {author} {\bibfnamefont {J.}~\bibnamefont {Wosnitza}}, \bibinfo
  {author} {\bibfnamefont {P.}~\bibnamefont {Lampen-Kelley}}, \bibinfo {author}
  {\bibfnamefont {A.}~\bibnamefont {Banerjee}}, \bibinfo {author}
  {\bibfnamefont {J.-Q.}\ \bibnamefont {Yan}}, \bibinfo {author} {\bibfnamefont
  {C.~A.}\ \bibnamefont {Bridges}}, \bibinfo {author} {\bibfnamefont {D.~G.}\
  \bibnamefont {Mandrus}}, \bibinfo {author} {\bibfnamefont {S.~E.}\
  \bibnamefont {Nagler}}, \bibinfo {author} {\bibfnamefont {A.~K.}\
  \bibnamefont {Kolezhuk}}, \ and\ \bibinfo {author} {\bibfnamefont {S.~A.}\
  \bibnamefont {Zvyagin}},\ }\href {\doibase 10.1103/PhysRevB.96.241107}
  {\bibfield  {journal} {\bibinfo  {journal} {Phys. Rev. B}\ }\textbf {\bibinfo
  {volume} {96}},\ \bibinfo {pages} {241107} (\bibinfo {year}
  {2017})}\BibitemShut {NoStop}%
\bibitem [{\citenamefont {Winter}\ \emph {et~al.}(2016)\citenamefont {Winter},
  \citenamefont {Li}, \citenamefont {Jeschke},\ and\ \citenamefont
  {Valent\'{\i}}}]{winter2016}%
  \BibitemOpen
  \bibfield  {author} {\bibinfo {author} {\bibfnamefont {S.~M.}\ \bibnamefont
  {Winter}}, \bibinfo {author} {\bibfnamefont {Y.}~\bibnamefont {Li}}, \bibinfo
  {author} {\bibfnamefont {H.~O.}\ \bibnamefont {Jeschke}}, \ and\ \bibinfo
  {author} {\bibfnamefont {R.}~\bibnamefont {Valent\'{\i}}},\ }\href {\doibase
  10.1103/PhysRevB.93.214431} {\bibfield  {journal} {\bibinfo  {journal} {Phys.
  Rev. B}\ }\textbf {\bibinfo {volume} {93}},\ \bibinfo {pages} {214431}
  (\bibinfo {year} {2016})}\BibitemShut {NoStop}%
\bibitem [{\citenamefont {Winter}\ \emph {et~al.}(2017)\citenamefont {Winter},
  \citenamefont {Riedl}, \citenamefont {Maksimov}, \citenamefont {Chernyshev},
  \citenamefont {Honecker},\ and\ \citenamefont
  {Valent{\'\i}}}]{winter:2017breakdown}%
  \BibitemOpen
  \bibfield  {author} {\bibinfo {author} {\bibfnamefont {S.~M.}\ \bibnamefont
  {Winter}}, \bibinfo {author} {\bibfnamefont {K.}~\bibnamefont {Riedl}},
  \bibinfo {author} {\bibfnamefont {P.~A.}\ \bibnamefont {Maksimov}}, \bibinfo
  {author} {\bibfnamefont {A.~L.}\ \bibnamefont {Chernyshev}}, \bibinfo
  {author} {\bibfnamefont {A.}~\bibnamefont {Honecker}}, \ and\ \bibinfo
  {author} {\bibfnamefont {R.}~\bibnamefont {Valent{\'\i}}},\ }\href@noop {}
  {\bibfield  {journal} {\bibinfo  {journal} {Nature Communications}\ }\textbf
  {\bibinfo {volume} {8}},\ \bibinfo {pages} {1152} (\bibinfo {year}
  {2017})}\BibitemShut {NoStop}%
\bibitem [{\citenamefont {Kim}\ and\ \citenamefont
  {Kee}(2016)}]{kim:2016crystal}%
  \BibitemOpen
  \bibfield  {author} {\bibinfo {author} {\bibfnamefont {H.-S.}\ \bibnamefont
  {Kim}}\ and\ \bibinfo {author} {\bibfnamefont {H.-Y.}\ \bibnamefont {Kee}},\
  }\href {\doibase 10.1103/PhysRevB.93.155143} {\bibfield  {journal} {\bibinfo
  {journal} {Phys. Rev. B}\ }\textbf {\bibinfo {volume} {93}},\ \bibinfo
  {pages} {155143} (\bibinfo {year} {2016})}\BibitemShut {NoStop}%
\bibitem [{\citenamefont {Kim}\ \emph {et~al.}(2015)\citenamefont {Kim},
  \citenamefont {V.}, \citenamefont {Catuneanu},\ and\ \citenamefont
  {Kee}}]{kim:2015kitaev}%
  \BibitemOpen
  \bibfield  {author} {\bibinfo {author} {\bibfnamefont {H.-S.}\ \bibnamefont
  {Kim}}, \bibinfo {author} {\bibfnamefont {V.~S.}\ \bibnamefont {V.}},
  \bibinfo {author} {\bibfnamefont {A.}~\bibnamefont {Catuneanu}}, \ and\
  \bibinfo {author} {\bibfnamefont {H.-Y.}\ \bibnamefont {Kee}},\ }\href
  {\doibase 10.1103/PhysRevB.91.241110} {\bibfield  {journal} {\bibinfo
  {journal} {Phys. Rev. B}\ }\textbf {\bibinfo {volume} {91}},\ \bibinfo
  {pages} {241110} (\bibinfo {year} {2015})}\BibitemShut {NoStop}%
\bibitem [{\citenamefont {Wang}\ \emph
  {et~al.}(2017{\natexlab{b}})\citenamefont {Wang}, \citenamefont {Dong},
  \citenamefont {Yu},\ and\ \citenamefont {Li}}]{wang:2017theoretical}%
  \BibitemOpen
  \bibfield  {author} {\bibinfo {author} {\bibfnamefont {W.}~\bibnamefont
  {Wang}}, \bibinfo {author} {\bibfnamefont {Z.-Y.}\ \bibnamefont {Dong}},
  \bibinfo {author} {\bibfnamefont {S.-L.}\ \bibnamefont {Yu}}, \ and\ \bibinfo
  {author} {\bibfnamefont {J.-X.}\ \bibnamefont {Li}},\ }\href {\doibase
  10.1103/PhysRevB.96.115103} {\bibfield  {journal} {\bibinfo  {journal} {Phys.
  Rev. B}\ }\textbf {\bibinfo {volume} {96}},\ \bibinfo {pages} {115103}
  (\bibinfo {year} {2017}{\natexlab{b}})}\BibitemShut {NoStop}%
\bibitem [{\citenamefont {Hou}\ \emph {et~al.}(2017)\citenamefont {Hou},
  \citenamefont {Xiang},\ and\ \citenamefont {Gong}}]{hou2017}%
  \BibitemOpen
  \bibfield  {author} {\bibinfo {author} {\bibfnamefont {Y.~S.}\ \bibnamefont
  {Hou}}, \bibinfo {author} {\bibfnamefont {H.~J.}\ \bibnamefont {Xiang}}, \
  and\ \bibinfo {author} {\bibfnamefont {X.~G.}\ \bibnamefont {Gong}},\ }\href
  {\doibase 10.1103/PhysRevB.96.054410} {\bibfield  {journal} {\bibinfo
  {journal} {Phys. Rev. B}\ }\textbf {\bibinfo {volume} {96}},\ \bibinfo
  {pages} {054410} (\bibinfo {year} {2017})}\BibitemShut {NoStop}%
\bibitem [{\citenamefont {Cao}\ \emph {et~al.}(2016)\citenamefont {Cao},
  \citenamefont {Banerjee}, \citenamefont {Yan}, \citenamefont {Bridges},
  \citenamefont {Lumsden}, \citenamefont {Mandrus}, \citenamefont {Tennant},
  \citenamefont {Chakoumakos},\ and\ \citenamefont {Nagler}}]{cao:2016low}%
  \BibitemOpen
  \bibfield  {author} {\bibinfo {author} {\bibfnamefont {H.~B.}\ \bibnamefont
  {Cao}}, \bibinfo {author} {\bibfnamefont {A.}~\bibnamefont {Banerjee}},
  \bibinfo {author} {\bibfnamefont {J.-Q.}\ \bibnamefont {Yan}}, \bibinfo
  {author} {\bibfnamefont {C.~A.}\ \bibnamefont {Bridges}}, \bibinfo {author}
  {\bibfnamefont {M.~D.}\ \bibnamefont {Lumsden}}, \bibinfo {author}
  {\bibfnamefont {D.~G.}\ \bibnamefont {Mandrus}}, \bibinfo {author}
  {\bibfnamefont {D.~A.}\ \bibnamefont {Tennant}}, \bibinfo {author}
  {\bibfnamefont {B.~C.}\ \bibnamefont {Chakoumakos}}, \ and\ \bibinfo {author}
  {\bibfnamefont {S.~E.}\ \bibnamefont {Nagler}},\ }\href {\doibase
  10.1103/PhysRevB.93.134423} {\bibfield  {journal} {\bibinfo  {journal} {Phys.
  Rev. B}\ }\textbf {\bibinfo {volume} {93}},\ \bibinfo {pages} {134423}
  (\bibinfo {year} {2016})}\BibitemShut {NoStop}%
\bibitem [{\citenamefont {Kubota}\ \emph {et~al.}(2015)\citenamefont {Kubota},
  \citenamefont {Tanaka}, \citenamefont {Ono}, \citenamefont {Narumi},\ and\
  \citenamefont {Kindo}}]{Kubota:2015Successive}%
  \BibitemOpen
  \bibfield  {author} {\bibinfo {author} {\bibfnamefont {Y.}~\bibnamefont
  {Kubota}}, \bibinfo {author} {\bibfnamefont {H.}~\bibnamefont {Tanaka}},
  \bibinfo {author} {\bibfnamefont {T.}~\bibnamefont {Ono}}, \bibinfo {author}
  {\bibfnamefont {Y.}~\bibnamefont {Narumi}}, \ and\ \bibinfo {author}
  {\bibfnamefont {K.}~\bibnamefont {Kindo}},\ }\href {\doibase
  10.1103/PhysRevB.91.094422} {\bibfield  {journal} {\bibinfo  {journal} {Phys.
  Rev. B}\ }\textbf {\bibinfo {volume} {91}},\ \bibinfo {pages} {094422}
  (\bibinfo {year} {2015})}\BibitemShut {NoStop}%
\bibitem [{\citenamefont {Nasu}\ \emph {et~al.}(2017)\citenamefont {Nasu},
  \citenamefont {Yoshitake},\ and\ \citenamefont {Motome}}]{nasu:2017thermal}%
  \BibitemOpen
  \bibfield  {author} {\bibinfo {author} {\bibfnamefont {J.}~\bibnamefont
  {Nasu}}, \bibinfo {author} {\bibfnamefont {J.}~\bibnamefont {Yoshitake}}, \
  and\ \bibinfo {author} {\bibfnamefont {Y.}~\bibnamefont {Motome}},\ }\href
  {\doibase 10.1103/PhysRevLett.119.127204} {\bibfield  {journal} {\bibinfo
  {journal} {Phys. Rev. Lett.}\ }\textbf {\bibinfo {volume} {119}},\ \bibinfo
  {pages} {127204} (\bibinfo {year} {2017})}\BibitemShut {NoStop}%
\bibitem [{\citenamefont {Kasahara}\ \emph
  {et~al.}(2018{\natexlab{b}})\citenamefont {Kasahara}, \citenamefont {Sugii},
  \citenamefont {Ohnishi}, \citenamefont {Shimozawa}, \citenamefont
  {Yamashita}, \citenamefont {Kurita}, \citenamefont {Tanaka}, \citenamefont
  {Nasu}, \citenamefont {Motome}, \citenamefont {Shibauchi},\ and\
  \citenamefont {Matsuda}}]{Kasahara2017}%
  \BibitemOpen
  \bibfield  {author} {\bibinfo {author} {\bibfnamefont {Y.}~\bibnamefont
  {Kasahara}}, \bibinfo {author} {\bibfnamefont {K.}~\bibnamefont {Sugii}},
  \bibinfo {author} {\bibfnamefont {T.}~\bibnamefont {Ohnishi}}, \bibinfo
  {author} {\bibfnamefont {M.}~\bibnamefont {Shimozawa}}, \bibinfo {author}
  {\bibfnamefont {M.}~\bibnamefont {Yamashita}}, \bibinfo {author}
  {\bibfnamefont {N.}~\bibnamefont {Kurita}}, \bibinfo {author} {\bibfnamefont
  {H.}~\bibnamefont {Tanaka}}, \bibinfo {author} {\bibfnamefont
  {J.}~\bibnamefont {Nasu}}, \bibinfo {author} {\bibfnamefont {Y.}~\bibnamefont
  {Motome}}, \bibinfo {author} {\bibfnamefont {T.}~\bibnamefont {Shibauchi}}, \
  and\ \bibinfo {author} {\bibfnamefont {Y.}~\bibnamefont {Matsuda}},\ }\href
  {\doibase 10.1103/PhysRevLett.120.217205} {\bibfield  {journal} {\bibinfo
  {journal} {Phys. Rev. Lett.}\ }\textbf {\bibinfo {volume} {120}},\ \bibinfo
  {pages} {217205} (\bibinfo {year} {2018}{\natexlab{b}})}\BibitemShut
  {NoStop}%
\bibitem [{\citenamefont {Hentrich}\ \emph {et~al.}(2018)\citenamefont
  {Hentrich}, \citenamefont {Roslova}, \citenamefont {Isaeva}, \citenamefont
  {Doert}, \citenamefont {Brenig}, \citenamefont {B{\"u}chner},\ and\
  \citenamefont {Hess}}]{hentrich:2018large}%
  \BibitemOpen
  \bibfield  {author} {\bibinfo {author} {\bibfnamefont {R.}~\bibnamefont
  {Hentrich}}, \bibinfo {author} {\bibfnamefont {M.}~\bibnamefont {Roslova}},
  \bibinfo {author} {\bibfnamefont {A.}~\bibnamefont {Isaeva}}, \bibinfo
  {author} {\bibfnamefont {T.}~\bibnamefont {Doert}}, \bibinfo {author}
  {\bibfnamefont {W.}~\bibnamefont {Brenig}}, \bibinfo {author} {\bibfnamefont
  {B.}~\bibnamefont {B{\"u}chner}}, \ and\ \bibinfo {author} {\bibfnamefont
  {C.}~\bibnamefont {Hess}},\ }\href@noop {} {\bibfield  {journal} {\bibinfo
  {journal} {arXiv preprint arXiv:1803.08162}\ } (\bibinfo {year}
  {2018})}\BibitemShut {NoStop}%
\bibitem [{\citenamefont {Vinkler-Aviv}\ and\ \citenamefont
  {Rosch}(2018)}]{vinkler:2018approximately}%
  \BibitemOpen
  \bibfield  {author} {\bibinfo {author} {\bibfnamefont {Y.}~\bibnamefont
  {Vinkler-Aviv}}\ and\ \bibinfo {author} {\bibfnamefont {A.}~\bibnamefont
  {Rosch}},\ }\href@noop {} {\bibfield  {journal} {\bibinfo  {journal} {arXiv
  preprint arXiv:1805.11587}\ } (\bibinfo {year} {2018})}\BibitemShut {NoStop}%
\bibitem [{\citenamefont {Ye}\ \emph {et~al.}(2018)\citenamefont {Ye},
  \citenamefont {Hal{\'a}sz}, \citenamefont {Savary},\ and\ \citenamefont
  {Balents}}]{ye:2018quantization}%
  \BibitemOpen
  \bibfield  {author} {\bibinfo {author} {\bibfnamefont {M.}~\bibnamefont
  {Ye}}, \bibinfo {author} {\bibfnamefont {G.~B.}\ \bibnamefont {Hal{\'a}sz}},
  \bibinfo {author} {\bibfnamefont {L.}~\bibnamefont {Savary}}, \ and\ \bibinfo
  {author} {\bibfnamefont {L.}~\bibnamefont {Balents}},\ }\href@noop {}
  {\bibfield  {journal} {\bibinfo  {journal} {arXiv preprint arXiv:1805.10532}\
  } (\bibinfo {year} {2018})}\BibitemShut {NoStop}%
\bibitem [{\citenamefont {Liu}\ and\ \citenamefont
  {Normand}(2018)}]{Liu:2018Dirac}%
  \BibitemOpen
  \bibfield  {author} {\bibinfo {author} {\bibfnamefont {Z.-X.}\ \bibnamefont
  {Liu}}\ and\ \bibinfo {author} {\bibfnamefont {B.}~\bibnamefont {Normand}},\
  }\href {\doibase 10.1103/PhysRevLett.120.187201} {\bibfield  {journal}
  {\bibinfo  {journal} {Phys. Rev. Lett.}\ }\textbf {\bibinfo {volume} {120}},\
  \bibinfo {pages} {187201} (\bibinfo {year} {2018})}\BibitemShut {NoStop}%
\bibitem [{\citenamefont {Inyushkin}\ and\ \citenamefont
  {Taldenkov}(2007)}]{Inyushkin:2007phonon}%
  \BibitemOpen
  \bibfield  {author} {\bibinfo {author} {\bibfnamefont {A.~V.}\ \bibnamefont
  {Inyushkin}}\ and\ \bibinfo {author} {\bibfnamefont {A.~N.}\ \bibnamefont
  {Taldenkov}},\ }\href {\doibase 10.1134/S0021364007180075} {\bibfield
  {journal} {\bibinfo  {journal} {JETP Letters}\ }\textbf {\bibinfo {volume}
  {86}},\ \bibinfo {pages} {379} (\bibinfo {year} {2007})}\BibitemShut
  {NoStop}%
\bibitem [{\citenamefont {Sugii}\ \emph {et~al.}(2017)\citenamefont {Sugii},
  \citenamefont {Shimozawa}, \citenamefont {Watanabe}, \citenamefont {Suzuki},
  \citenamefont {Halim}, \citenamefont {Kimata}, \citenamefont {Matsumoto},
  \citenamefont {Nakatsuji},\ and\ \citenamefont
  {Yamashita}}]{Sugii:2017phonon}%
  \BibitemOpen
  \bibfield  {author} {\bibinfo {author} {\bibfnamefont {K.}~\bibnamefont
  {Sugii}}, \bibinfo {author} {\bibfnamefont {M.}~\bibnamefont {Shimozawa}},
  \bibinfo {author} {\bibfnamefont {D.}~\bibnamefont {Watanabe}}, \bibinfo
  {author} {\bibfnamefont {Y.}~\bibnamefont {Suzuki}}, \bibinfo {author}
  {\bibfnamefont {M.}~\bibnamefont {Halim}}, \bibinfo {author} {\bibfnamefont
  {M.}~\bibnamefont {Kimata}}, \bibinfo {author} {\bibfnamefont
  {Y.}~\bibnamefont {Matsumoto}}, \bibinfo {author} {\bibfnamefont
  {S.}~\bibnamefont {Nakatsuji}}, \ and\ \bibinfo {author} {\bibfnamefont
  {M.}~\bibnamefont {Yamashita}},\ }\href {\doibase
  10.1103/PhysRevLett.118.145902} {\bibfield  {journal} {\bibinfo  {journal}
  {Phys. Rev. Lett.}\ }\textbf {\bibinfo {volume} {118}},\ \bibinfo {pages}
  {145902} (\bibinfo {year} {2017})}\BibitemShut {NoStop}%
\bibitem [{\citenamefont {Matsumoto}\ \emph {et~al.}(2014)\citenamefont
  {Matsumoto}, \citenamefont {Shindou},\ and\ \citenamefont
  {Murakami}}]{matsumoto2014}%
  \BibitemOpen
  \bibfield  {author} {\bibinfo {author} {\bibfnamefont {R.}~\bibnamefont
  {Matsumoto}}, \bibinfo {author} {\bibfnamefont {R.}~\bibnamefont {Shindou}},
  \ and\ \bibinfo {author} {\bibfnamefont {S.}~\bibnamefont {Murakami}},\
  }\href {\doibase 10.1103/PhysRevB.89.054420} {\bibfield  {journal} {\bibinfo
  {journal} {Phys. Rev. B}\ }\textbf {\bibinfo {volume} {89}},\ \bibinfo
  {pages} {054420} (\bibinfo {year} {2014})}\BibitemShut {NoStop}%
\bibitem [{\citenamefont {Hoon~Lee}\ \emph {et~al.}(2017)\citenamefont
  {Hoon~Lee}, \citenamefont {Chung}, \citenamefont {Park},\ and\ \citenamefont
  {Park}}]{lee2017}%
  \BibitemOpen
  \bibfield  {author} {\bibinfo {author} {\bibfnamefont {K.}~\bibnamefont
  {Hoon~Lee}}, \bibinfo {author} {\bibfnamefont {S.}~\bibnamefont {Chung}},
  \bibinfo {author} {\bibfnamefont {K.}~\bibnamefont {Park}}, \ and\ \bibinfo
  {author} {\bibfnamefont {J.-G.}\ \bibnamefont {Park}},\ }\href@noop {} {\
  (\bibinfo {year} {2017})}\BibitemShut {NoStop}%
\bibitem [{\citenamefont
  {Owerre}(2016{\natexlab{a}})}]{owerre:2016topological}%
  \BibitemOpen
  \bibfield  {author} {\bibinfo {author} {\bibfnamefont {S.~A.}\ \bibnamefont
  {Owerre}},\ }\href {\doibase 10.1063/1.4959815} {\bibfield  {journal}
  {\bibinfo  {journal} {Journal of Applied Physics}\ }\textbf {\bibinfo
  {volume} {120}},\ \bibinfo {pages} {043903} (\bibinfo {year}
  {2016}{\natexlab{a}})},\ \Eprint
  {http://arxiv.org/abs/https://doi.org/10.1063/1.4959815}
  {https://doi.org/10.1063/1.4959815} \BibitemShut {NoStop}%
\bibitem [{\citenamefont {Owerre}(2016{\natexlab{b}})}]{owerre2016magnon}%
  \BibitemOpen
  \bibfield  {author} {\bibinfo {author} {\bibfnamefont {S.~A.}\ \bibnamefont
  {Owerre}},\ }\href {\doibase 10.1103/PhysRevB.94.094405} {\bibfield
  {journal} {\bibinfo  {journal} {Phys. Rev. B}\ }\textbf {\bibinfo {volume}
  {94}},\ \bibinfo {pages} {094405} (\bibinfo {year}
  {2016}{\natexlab{b}})}\BibitemShut {NoStop}%
\bibitem [{\citenamefont {McClarty}\ \emph {et~al.}(2018)\citenamefont
  {McClarty}, \citenamefont {Dong}, \citenamefont {Gohlke}, \citenamefont
  {Rau}, \citenamefont {Pollmann}, \citenamefont {Moessner},\ and\
  \citenamefont {Penc}}]{mcclarty:2018topological}%
  \BibitemOpen
  \bibfield  {author} {\bibinfo {author} {\bibfnamefont {P.}~\bibnamefont
  {McClarty}}, \bibinfo {author} {\bibfnamefont {X.-Y.}\ \bibnamefont {Dong}},
  \bibinfo {author} {\bibfnamefont {M.}~\bibnamefont {Gohlke}}, \bibinfo
  {author} {\bibfnamefont {J.}~\bibnamefont {Rau}}, \bibinfo {author}
  {\bibfnamefont {F.}~\bibnamefont {Pollmann}}, \bibinfo {author}
  {\bibfnamefont {R.}~\bibnamefont {Moessner}}, \ and\ \bibinfo {author}
  {\bibfnamefont {K.}~\bibnamefont {Penc}},\ }\href@noop {} {\bibfield
  {journal} {\bibinfo  {journal} {arXiv preprint arXiv:1802.04283}\ } (\bibinfo
  {year} {2018})}\BibitemShut {NoStop}%
\bibitem [{\citenamefont {Jones}\ \emph {et~al.}(1987)\citenamefont {Jones},
  \citenamefont {Pankhurst},\ and\ \citenamefont {Johnson}}]{Jones}%
  \BibitemOpen
  \bibfield  {author} {\bibinfo {author} {\bibfnamefont {D.~H.}\ \bibnamefont
  {Jones}}, \bibinfo {author} {\bibfnamefont {Q.~A.}\ \bibnamefont
  {Pankhurst}}, \ and\ \bibinfo {author} {\bibfnamefont {C.~E.}\ \bibnamefont
  {Johnson}},\ }\href {http://stacks.iop.org/0022-3719/20/i=31/a=017}
  {\bibfield  {journal} {\bibinfo  {journal} {Journal of Physics C: Solid State
  Physics}\ }\textbf {\bibinfo {volume} {20}},\ \bibinfo {pages} {5149}
  (\bibinfo {year} {1987})}\BibitemShut {NoStop}%
\bibitem [{\citenamefont {Yadav}\ \emph {et~al.}(2016)\citenamefont {Yadav},
  \citenamefont {Bogdanov}, \citenamefont {Katukuri}, \citenamefont
  {Nishimoto}, \citenamefont {Van Den~Brink},\ and\ \citenamefont
  {Hozoi}}]{yadav:2016kitaev}%
  \BibitemOpen
  \bibfield  {author} {\bibinfo {author} {\bibfnamefont {R.}~\bibnamefont
  {Yadav}}, \bibinfo {author} {\bibfnamefont {N.~A.}\ \bibnamefont {Bogdanov}},
  \bibinfo {author} {\bibfnamefont {V.~M.}\ \bibnamefont {Katukuri}}, \bibinfo
  {author} {\bibfnamefont {S.}~\bibnamefont {Nishimoto}}, \bibinfo {author}
  {\bibfnamefont {J.}~\bibnamefont {Van Den~Brink}}, \ and\ \bibinfo {author}
  {\bibfnamefont {L.}~\bibnamefont {Hozoi}},\ }\href@noop {} {\bibfield
  {journal} {\bibinfo  {journal} {Scientific reports}\ }\textbf {\bibinfo
  {volume} {6}},\ \bibinfo {pages} {37925} (\bibinfo {year}
  {2016})}\BibitemShut {NoStop}%
\bibitem [{\citenamefont {Sears}\ \emph {et~al.}(2015)\citenamefont {Sears},
  \citenamefont {Songvilay}, \citenamefont {Plumb}, \citenamefont {Clancy},
  \citenamefont {Qiu}, \citenamefont {Zhao}, \citenamefont {Parshall},\ and\
  \citenamefont {Kim}}]{sears:2015moment}%
  \BibitemOpen
  \bibfield  {author} {\bibinfo {author} {\bibfnamefont {J.~A.}\ \bibnamefont
  {Sears}}, \bibinfo {author} {\bibfnamefont {M.}~\bibnamefont {Songvilay}},
  \bibinfo {author} {\bibfnamefont {K.~W.}\ \bibnamefont {Plumb}}, \bibinfo
  {author} {\bibfnamefont {J.~P.}\ \bibnamefont {Clancy}}, \bibinfo {author}
  {\bibfnamefont {Y.}~\bibnamefont {Qiu}}, \bibinfo {author} {\bibfnamefont
  {Y.}~\bibnamefont {Zhao}}, \bibinfo {author} {\bibfnamefont {D.}~\bibnamefont
  {Parshall}}, \ and\ \bibinfo {author} {\bibfnamefont {Y.-J.}\ \bibnamefont
  {Kim}},\ }\href {\doibase 10.1103/PhysRevB.91.144420} {\bibfield  {journal}
  {\bibinfo  {journal} {Phys. Rev. B}\ }\textbf {\bibinfo {volume} {91}},\
  \bibinfo {pages} {144420} (\bibinfo {year} {2015})}\BibitemShut {NoStop}%
\bibitem [{\citenamefont {Majumder}\ \emph {et~al.}(2015)\citenamefont
  {Majumder}, \citenamefont {Schmidt}, \citenamefont {Rosner}, \citenamefont
  {Tsirlin}, \citenamefont {Yasuoka},\ and\ \citenamefont
  {Baenitz}}]{majumder:2015anisotropic}%
  \BibitemOpen
  \bibfield  {author} {\bibinfo {author} {\bibfnamefont {M.}~\bibnamefont
  {Majumder}}, \bibinfo {author} {\bibfnamefont {M.}~\bibnamefont {Schmidt}},
  \bibinfo {author} {\bibfnamefont {H.}~\bibnamefont {Rosner}}, \bibinfo
  {author} {\bibfnamefont {A.~A.}\ \bibnamefont {Tsirlin}}, \bibinfo {author}
  {\bibfnamefont {H.}~\bibnamefont {Yasuoka}}, \ and\ \bibinfo {author}
  {\bibfnamefont {M.}~\bibnamefont {Baenitz}},\ }\href {\doibase
  10.1103/PhysRevB.91.180401} {\bibfield  {journal} {\bibinfo  {journal} {Phys.
  Rev. B}\ }\textbf {\bibinfo {volume} {91}},\ \bibinfo {pages} {180401}
  (\bibinfo {year} {2015})}\BibitemShut {NoStop}%
\bibitem [{\citenamefont {Johnson}\ \emph {et~al.}(2015)\citenamefont
  {Johnson}, \citenamefont {Williams}, \citenamefont {Haghighirad},
  \citenamefont {Singleton}, \citenamefont {Zapf}, \citenamefont {Manuel},
  \citenamefont {Mazin}, \citenamefont {Li}, \citenamefont {Jeschke},
  \citenamefont {Valent\'{\i}},\ and\ \citenamefont
  {Coldea}}]{johnson:2015monoclinic}%
  \BibitemOpen
  \bibfield  {author} {\bibinfo {author} {\bibfnamefont {R.~D.}\ \bibnamefont
  {Johnson}}, \bibinfo {author} {\bibfnamefont {S.~C.}\ \bibnamefont
  {Williams}}, \bibinfo {author} {\bibfnamefont {A.~A.}\ \bibnamefont
  {Haghighirad}}, \bibinfo {author} {\bibfnamefont {J.}~\bibnamefont
  {Singleton}}, \bibinfo {author} {\bibfnamefont {V.}~\bibnamefont {Zapf}},
  \bibinfo {author} {\bibfnamefont {P.}~\bibnamefont {Manuel}}, \bibinfo
  {author} {\bibfnamefont {I.~I.}\ \bibnamefont {Mazin}}, \bibinfo {author}
  {\bibfnamefont {Y.}~\bibnamefont {Li}}, \bibinfo {author} {\bibfnamefont
  {H.~O.}\ \bibnamefont {Jeschke}}, \bibinfo {author} {\bibfnamefont
  {R.}~\bibnamefont {Valent\'{\i}}}, \ and\ \bibinfo {author} {\bibfnamefont
  {R.}~\bibnamefont {Coldea}},\ }\href {\doibase 10.1103/PhysRevB.92.235119}
  {\bibfield  {journal} {\bibinfo  {journal} {Phys. Rev. B}\ }\textbf {\bibinfo
  {volume} {92}},\ \bibinfo {pages} {235119} (\bibinfo {year}
  {2015})}\BibitemShut {NoStop}%
\bibitem [{\citenamefont {Janssen}\ \emph {et~al.}(2017)\citenamefont
  {Janssen}, \citenamefont {Andrade},\ and\ \citenamefont {Vojta}}]{paramsum}%
  \BibitemOpen
  \bibfield  {author} {\bibinfo {author} {\bibfnamefont {L.}~\bibnamefont
  {Janssen}}, \bibinfo {author} {\bibfnamefont {E.~C.}\ \bibnamefont
  {Andrade}}, \ and\ \bibinfo {author} {\bibfnamefont {M.}~\bibnamefont
  {Vojta}},\ }\href {\doibase 10.1103/PhysRevB.96.064430} {\bibfield  {journal}
  {\bibinfo  {journal} {Phys. Rev. B}\ }\textbf {\bibinfo {volume} {96}},\
  \bibinfo {pages} {064430} (\bibinfo {year} {2017})}\BibitemShut {NoStop}%
\bibitem [{\citenamefont {Hirobe}\ \emph {et~al.}(2017)\citenamefont {Hirobe},
  \citenamefont {Sato}, \citenamefont {Shiomi}, \citenamefont {Tanaka},\ and\
  \citenamefont {Saitoh}}]{hirobe:2017magnetic}%
  \BibitemOpen
  \bibfield  {author} {\bibinfo {author} {\bibfnamefont {D.}~\bibnamefont
  {Hirobe}}, \bibinfo {author} {\bibfnamefont {M.}~\bibnamefont {Sato}},
  \bibinfo {author} {\bibfnamefont {Y.}~\bibnamefont {Shiomi}}, \bibinfo
  {author} {\bibfnamefont {H.}~\bibnamefont {Tanaka}}, \ and\ \bibinfo {author}
  {\bibfnamefont {E.}~\bibnamefont {Saitoh}},\ }\href {\doibase
  10.1103/PhysRevB.95.241112} {\bibfield  {journal} {\bibinfo  {journal} {Phys.
  Rev. B}\ }\textbf {\bibinfo {volume} {95}},\ \bibinfo {pages} {241112}
  (\bibinfo {year} {2017})}\BibitemShut {NoStop}%
\end{thebibliography}%

\clearpage

\section*{Supplemental Material: Derivation of SWT Hamiltonian}\label{SWTderiv}

See Fig.~\ref{fig:conventions} for conventions. We essentially follow Ref.~\cite{Jones}. We start with Eq.~\eqref{eq:Ham}, and define the rotated spin as $\Omega$ with 
\begin{equation}
\vec S = R \vec \Omega =
\begin{scriptsize}
  	\begin{bmatrix}
  	\cos(\theta)\cos(\phi) & -\sin(\phi) & \sin(\theta)\cos(\phi)\\
  	\cos(\theta)\sin(\phi) & \cos(\phi) & \sin(\theta)\sin(\phi) \\
  	-\sin(\theta) & 0 & \cos(\theta)
  	\end{bmatrix}	\vec \Omega 
\end{scriptsize}
\end{equation}
where $(\theta,\phi)$ specify the spin direction in polar coordinates. 
We will write $S_{j}^X = R_X \Omega_{j}^X$ for $i \in \{A,B,C,D\}$ and $j$ indicating the lattice point $r_j$ to indicate that we are rotating the four sublattices potentially differently, with $R_X = R(\theta_X,\phi_X)$. Let $X(j)$ denote the sublattice that $r_j$ belongs to.

Defining
\begin{equation}
\begin{aligned}
M^q_{XY} = R_X^T \cdot &(J_1 + T^q) \cdot R_Y;\qquad N_{XY} = J_3 R_X^T R_Y \\
T^x = \begin{scriptsize} \begin{bmatrix} K &\Gamma' & \Gamma'\\ \Gamma' & 0 & \Gamma \\ \Gamma'& \Gamma & 0 \end{bmatrix}\end{scriptsize};&\ \ \ 
T^y= \begin{scriptsize}\begin{bmatrix}  0& \Gamma'& \Gamma \\ \Gamma'& K &\Gamma'  \\ \Gamma & \Gamma' & 0\end{bmatrix}\end{scriptsize};\ \ \ 
T^z = \begin{scriptsize}\begin{bmatrix}  0 & \Gamma & \Gamma' \\ \Gamma & 0 &\Gamma' \\ \Gamma'& \Gamma'& K \end{bmatrix}\end{scriptsize};
\end{aligned}
\end{equation}
we will rewrite the Hamiltonian as follows:
\begin{equation}
\begin{aligned}
H&=\sum_{j: X(j) \in {A,C}}  \sum_{q \in \{x,y,z\}}\left( {\bf \Omega}_{j}^X \cdot M^q_{XY} \cdot {\bf \Omega}_{j+\alpha_q}^Y \right.\\
+&\left.  {\bf \Omega}_{j}^X \cdot N_{XZ} \cdot {\bf \Omega}_{j-2\alpha_q}^Z\right)
- \sum_{j:  X(j) \in \{A,B,C,D\}} {\bf h} \cdot R_X \cdot{\bf \Omega}_{j}^X,
\end{aligned}
\end{equation}
with $(X,Y,Z) = (X(j),X(j+\alpha_q),X(j-2\alpha_q))$, and where we have defined $\alpha_q$ as the vector going along a $q$ bond---$\alpha_x=(0,-1,1)/\sqrt{2}, \alpha_y = (1,0,-1)/\sqrt{2},$ and $\alpha_z=(-1,1,0)/\sqrt{2}$. 

Now, we switch to Holstein-Primakov (HP) bosons:
$\Omega_{j,x}^i \approx (S/2)^{1/2}(b_j^{i,\dagger} + b_{j}^i), \Omega_{j,y}^i \approx i (S/2)^{1/2}(b_j^{i,\dagger}-b_j^i)$, and $\Omega_{j,z}^i = S - b_{j}^{i,\dagger}b_{j}^i$
where $b_{j}^i$ is the boson annhilation operator at position $x_j$ on the sublattice $i$. This transformation is valid so long as the spin reduction is small, or , equivalently $\langle b_j^{i,\dagger}b_j^i \rangle / S \ll 1$.

We obtain a Hamiltonian of the form $H=H_0 + H_1 + H_2 + ...$ where $H_i$ has terms with $i$ bosons in them. The overall constant is unimportant. The condition that $H_1=0$ fixes the spin direction, and we find the following eight equations must hold (fixing the variables $(\theta_X,\phi_X)$ for $X\in \{A,B,C,D\}$).:
\begin{equation}
\begin{footnotesize}
\begin{aligned}
 0&=\sum_{q \in \{x,y\}} (M_{AB,m3}^q) + M_{AD,m3}^z +3N_{AD,m3} - ({\bf h}\cdot R_A)_m/S  \\
0&= \sum_{q \in \{x,y\}} (M_{AB,3m}^q) + M_{CB,3m}^z +3N_{CB,3m} - ({\bf h}\cdot R_B)_m/S  \\
&\qquad \text{ ($A\leftrightarrow C, B\leftrightarrow D$)}
\end{aligned}
\end{footnotesize}
\end{equation}
with $m\in \{1,2\}$.

Since $(\theta_X,\phi_X)$ are now fixed, we can find $H_2$.
Defining

\begin{equation}
\begin{aligned}
Q^q_{XY} &= \begin{scriptsize}\frac14\begin{bmatrix} 1 & -i \\ 1 & i \end{bmatrix} \begin{bmatrix} M_{XY,11}^q & M_{XY,12}^q \\ M_{XY,21}^q & M_{XY,22}^q \end{bmatrix} \begin{bmatrix} 1 & 1\\ -i &  i \end{bmatrix}\end{scriptsize};\\
P_{XY}&=\begin{scriptsize}\frac14\begin{bmatrix} 1 & -i \\ 1 & i \end{bmatrix} \begin{bmatrix} N_{XY,11} & N_{XY,12}\\ N_{XY,21}& N_{XY,22}\end{bmatrix} \begin{bmatrix} 1 & 1 \\ -i & i \end{bmatrix}\end{scriptsize},
\end{aligned}
\end{equation}
the expression for
$H_2$ will have many terms of the form $b_{j}^{X,\cdot/\dagger}b_{j+\alpha}^{X,\cdot/\dagger}$ (where $\cdot$ is the annihilation operator). Using the convention $
b_{k}^i = \frac{1}{\sqrt{N}}\sum_{X(j)\in i} b_{j}^i e^{- i \vec k \cdot \vec x_i}$ (with $N$ the number of points in one of the four sublattices),
we arrive at the Fourier transformed $H_2$ in the form
\begin{equation*}
H_2 = \frac12 \sum_k \psi_k^\dagger \mathcal H \psi_k
=\frac12 \sum_k \psi_k^\dagger \frac{S}{2}\begin{bmatrix} \mathcal A(k) & \mathcal B(k) \\ \mathcal B(-k)^* & \mathcal A(-k)^T\end{bmatrix} \psi_k
\end{equation*}
with $\psi_k^\dagger = (b_k^{A,\dagger},b_k^{B,\dagger},b_k^{C,\dagger},b_k^{D,\dagger},b_{-k}^A,b_{-k}^B,b_{-k}^C,b_{-k}^D)$. To do so, we averaged the $k\leftrightarrow -k$ expression, dividing by 2, and dropping a constant term. In carrying out the algebra, we find
\begin{equation*}
\begin{scriptsize}
\begin{bmatrix} \mathcal A(k) & \mathcal B(k) \end{bmatrix}= \begin{bmatrix}  C_{AA}^{\dagger \cdot}& C_{AB}^{\dagger \cdot}& 0 & C_{AD}^{\dagger \cdot} & 0 &\tilde C_{AB}^{\dagger \dagger} & 0 & \tilde C_{AD}^{\dagger\dagger} \\ 
C_{AB}^{\cdot \dagger} & C_{BB}^{\dagger \cdot} & C_{CB}^{\cdot \dagger} & 0 & C_{AB}^{\dagger\dagger} & 0 & C_{CB}^{\dagger \dagger} & 0 \\
0 & C_{CB}^{\dagger \cdot} & C_{CC}^{\dagger \cdot} &C_{CD}^{\dagger \cdot} &0 &\tilde C_{CB}^{\dagger\dagger} & 0& \tilde C_{CD}^{\dagger \dagger}\\
C_{AD}^{\cdot \dagger}& 0& C_{CD}^{\cdot \dagger} & C_{DD}^{\dagger \cdot} & C_{AD}^{\dagger\dagger} &0 &C_{CD}^{\dagger\dagger} &0 \\\end{bmatrix} 
\end{scriptsize}
\end{equation*}
where we have defined
\begin{equation}
\begin{footnotesize}
\begin{aligned}
C_{AA}^{\dagger \cdot} &=-2\sum_{q \in \{x,y\}} M^q_{AB,33}-2M^z_{AD,33}-6N_{AD,33} + \frac{2}{S}
({\bf h}\cdot R_A)_3 \\
C_{BB}^{\dagger \cdot}&=-2\sum_{q \in \{x,y\}} M^q_{AB,33}-2M^z_{CB,33}-6N_{CB,33}+\frac{2}{S}
({\bf h}\cdot R_B)_3 \\
C_{AB}^{\dagger \cdot}& =(C_{AB}^{\cdot \dagger})^* = \sum_{q \in \{x,y\}} 4Q_{AB,21}^qe^{+i\vec k \cdot \alpha_q}\\
C_{AB}^{\dagger \dagger}& = (C_{AB}^{\cdot \cdot})^*  =  \sum_{q \in \{x,y\}} 4Q_{AB,22}^qe^{-i\vec k \cdot \alpha_q}\\
C_{AD}^{\dagger \cdot}& = (C_{AD}^{\cdot \dagger})^* = 4Q_{AD,21}^z e^{i\vec k \cdot\alpha_z}
+4\sum_{q \in \{x,y,z\}} P_{AD,21} e^{-2 i\vec k \cdot\alpha_q}\\
C_{AD}^{\dagger \dagger} &= (C_{AD}^{\cdot \cdot})^* = 4Q_{AD,22}^z e^{-i\vec k \cdot\alpha_z}
+4\sum_{q \in \{x,y,z\}} P_{AD,22} e^{2 i\vec k \cdot\alpha_q}\\
&(A\leftrightarrow C, B\leftrightarrow D)
\end{aligned}
\end{footnotesize}
\end{equation}
with $\tilde C_{XY}^{\cdot/\dagger,\cdot/\dagger}(k) = C_{XY}^{\cdot/\dagger,\cdot/\dagger}(-k)$. The band energies $\omega_n(k)$ and Bogoliubov transformation matrix, $T_k$ are, respectively, the eigenvalues and the matrix of the eigenvectors of $\sigma_3 \mathcal H$.

\end{document}